\documentclass[useAMS,usenatbib]{mn2e}
\usepackage{graphicx}
\usepackage[space]{grffile}
\usepackage{latexsym}
\usepackage{natbib}
\usepackage{url}
\usepackage{color}
\usepackage[utf8]{inputenc}
\usepackage{longtable,lscape}
\usepackage{fancyref}
\usepackage{times}
\usepackage{amssymb}

\usepackage{color}
\usepackage{soul}

\def\kms{$\mbox{km s}^{-1}$}

\newcommand{\msun}{{\rm M}_\odot}

\long\def\symbolfootnote[#1]#2{\begingroup%
\def\thefootnote{\fnsymbol{footnote}}\footnote[#1]{#2}\endgroup} 
\def\aj{AJ}%          % Astronomical Journal
\def\araa{ARA\&A}%          % Annual Review of Astron and Astrophys
\def\apj{ApJ}%          % Astrophysical Journal
\def\apjl{ApJ}%          % Astrophysical Journal, Letters
\def\apjs{ApJS}%          % Astrophysical Journal, Supplement
\def\apss{Ap\&SS}%          % Astrophysics and Space Science
\def\aap{A\&A}%          % Astronomy and Astrophysics
\def\aapr{A\&A~Rev.}%          % Astronomy and Astrophysics Reviews
\def\mnras{MNRAS}%          % Monthly Notices of the RAS
\def\nar{New A Rev.}%          % New Astronomy Review
\def\pasp{PASP}%          % Publications of the ASP
\def\sovast{Soviet Ast.}

\usepackage{txfonts}
\def\xvec{\textbf{\emph{x}}}

\title[A faster rotation for the ``extreme" chemical subpopulation in M13]{Differences in the rotational properties of multiple stellar populations in M\,13: a faster rotation for the ``extreme" chemical subpopulation}
\author[Cordero et al.]{M. J. Cordero$^1$\thanks{E-mail: mjcorde@ari.uni-heidelberg.de}\footnotemark[3], V. H\'{e}nault-Brunet$^{2,3}$\thanks{E-mail: v.henault-brunet@astro.ru.nl}\thanks{The two lead authors are listed alphabetically.}, C. A. Pilachowski$^4$, E. Balbinot$^3$,  \newauthor C. I. Johnson$^{5}$, A. L. Varri$^6$\\
$^1$ Astronomisches Rechen-Institut, Zentrum f\"ur Astronomie der Universit\"at Heidelberg, M\"onchhofstrasse 12-14, D-69120 Heidelberg, Germany\\
$^2$ Department of Astrophysics/IMAPP, Radboud University, PO Box 9010, NL-6500 GL Nijmegen, the Netherlands\\
$^3$ Department of Physics, Faculty of Engineering and Physical Sciences, University of Surrey, Guildford, GU2 7XH, UK\\
$^4$ Astronomy Department, Indiana University Bloomington, Swain West 319, 727 East 3rd Street, Bloomington, IN 47405-7105\\
$^5$Harvard-Smithsonian Center for Astrophysics, 60 Garden Street, MS-15, Cambridge, MA 02138, USA\\
$^6$Institute for Astronomy, University of Edinburgh, Royal Observatory, Blackford Hill, Edinburgh EH9 3HJ, UK\\
}

\begin{document}

\date{Accepted ---. Received --- ; in original form ---}

\pagerange{\pageref{firstpage}--21} \pubyear{2016}

\maketitle

\label{firstpage}

\begin{abstract}
We use radial velocities from spectra of giants obtained with the WIYN telescope, coupled with existing chemical abundance measurements of Na and O for the same stars, to probe the presence of kinematic differences among the multiple populations of the globular cluster (GC) M13. To characterise the kinematics of various chemical subsamples, we introduce a method using Bayesian inference along with an MCMC algorithm to fit a six-parameter kinematic model (including rotation) to these subsamples. We find that the so-called ``extreme" population (Na-enhanced and extremely O-depleted) exhibits faster rotation around the centre of the cluster than the other cluster stars, in particular when compared to the dominant ``intermediate" population (moderately Na-enhanced and O-depleted). The most likely difference between the rotational amplitude of this extreme population and that of the intermediate population is found to be $\sim4$~\kms , with a 98.4\% probability that the rotational amplitude of the extreme population is larger than that of the intermediate population. We argue that the observed difference in rotational amplitudes, obtained when splitting subsamples according to their chemistry, is not a product of the long-term dynamical evolution of the cluster, but more likely a surviving feature imprinted early in the formation history of this GC and its multiple populations. We also find an agreement (within uncertainties) in the inferred position angle of the rotation axis of the different subpopulations considered. We discuss the constraints that these results may place on various formation scenarios.

\end{abstract}


\section{Introduction}
\label{Intro}

All well-studied Galactic globular clusters (GCs) exhibit star-to-star abundance variations with enhancement in N, Na, and Al, and depletion in C and O \citep[and occasionally depletion in Mg; e.g.][]{Gratton2012}. The study of \citet{Carretta2009} has revealed that typically $60-70\%$ of the cluster stars show light-element abundance variations that are not usually found in field stars, although recent work by \citet{Milone2016} showed that the fractions of cluster stars with chemical composition differing from halo stars are less uniform (ranging from $\sim$33\% to $\sim$92\%) than those from the sample of \citet{Carretta2009}\footnote{Note that the observed fraction of enriched stars depends on the radial distribution of the subpopulations and the region of the cluster sampled, and may thus differ from the true ``global" fraction of enriched stars. \citet{Milone2016} measure the ``local" fraction in the central regions of the clusters from Hubble Space Telescope (HST) photometry, while \citet{Carretta2009} infer it from spectroscopic studies sampling more external regions.}. Understanding the origin of the light element abundance spreads among GCs is one of the great unsolved problems in the study of stellar populations, and its solution potentially encodes crucial information that would reveal how GCs form.

Two popular scenarios for the formation of multiple populations invoke multiple star formation episodes, where later generations form out of gas polluted by the ejecta of an earlier generation of stars. These scenarios typically adopt fast rotating massive stars \citep[e.g.][]{Decressin2007} or asymptotic giant branch stars \citep[AGB; e.g.][]{DErcole2008} as the stellar candidates responsible for the chemical anomalies observed ubiquitously in GCs\footnote{One possible exception is Ruprecht 106, a relatively young and massive GC which appears to have a homogeneous chemical composition \citep{Villanova2013}, although it is currently unclear whether the apparent homogeneity could be due to a preferential sampling of ``primordial" stars towards the outskirts of this cluster and/or to uncertainties/biases in the reported Na abundances stemming from their adopted NLTE corrections. Furthermore, it is worth noting that the low abundances of alpha elements in Rup 106 (with respect to Galactic GCs) suggest an accreted origin. The low-mass Galactic GC E~3 also displays a CN abundance pattern consistent with a single stellar population with no self-enrichment \citep{SS2015}.}. In both of these scenarios, the mass of stellar ejecta available to form later generations is insufficient to account for the large observed fractions of enriched stars. To overcome this issue, these scenarios either adopt a non-standard initial mass function for the different generations \citep[e.g.][]{Decressin2007} and/or predict that GCs were 10-100 times more massive at birth \citep[e.g.][]{DErcole2008}. This last implication is however in conflict with observations of the number ratio of GC and field stars in nearby dwarf galaxies (Fornax, WLM, IKN), which suggest that GCs in these systems could not have been more than about five times more massive initially \citep{Larsen2012, Larsen2014}. In addition to the two sources of enriched material mentioned above, super-massive stars with $M\sim10^4 \ \msun$ have also been considered as polluters to explain the detailed abundance patterns observed in GCs \citep{Denissenkov2014, Denissenkov2015}. \citet{Bastian2013} also proposed a scenario that does not invoke multiple generations of stars, and in which the enriched material is released by massive interacting binaries \citep{deMink2009} and accreted onto the circumstellar discs of low-mass pre-main sequence stars from the same stellar generation \citep[although see][which highlights important limitations regarding the efficiency of such an accretion process]{Wijnen2016}. For a more detailed review of the different formation scenarios for multiple populations developed so far, in particular with respect to their shortcomings in simultaneously reproducing the observed abundance patterns of Na, O, and He in different GCs, we refer to \citet{Bastian2015}, \citet{Bastianetal2015}, and \citet{Renzini2015}. In short, none of the scenarios proposed so far provides a complete and satisfying explanation to the presence of multiple populations in GCs.

Understanding the formation and evolution of GCs and their multiple populations requires theoretical and observational efforts. From the observational perspective, the presence of light-element abundance variations has been probed by using low- \citep[e.g.][]{Smith1996, Harbeck2003, Smolinski2011} or high-resolution spectroscopy \citep[e.g.][]{Kraft1994, Gratton2004, Carretta2009}, and with photometry using narrow band filters covering molecular bands sensitive to the abundances of C, N, and O abundances \citep[e.g.][]{Lardo2011, Piotto2015, Massari2016}, where the light-element abundance spreads show up as split or broad sequences in the colour-magnitude diagrams (CMDs).

Additional insights have emerged by studying large samples of stars covering a wide spatial range in several GCs. For instance, \citet{Lardo2011} found a difference in the radial distributions (from about the half-light radius and beyond) of the different populations identified using Sloan $u-g$ photometry in a number of GCs (M2, M3, M5, M13, M15, M53, and M92). In their study, the presumably N-enriched stars were found to be more centrally concentrated. Previous low-resolution spectroscopy studies have also found that the N-rich population is more centrally concentrated in the metal-rich GC 47 Tuc \citep[e.g.][]{Briley1997}. This finding was confirmed by \citet{Milone2012} using HST photometry and \citet{Cordero2014} using high-resolution spectroscopy. A spatial separation among the multiple populations in GCs could have been imprinted during the early stages of GC formation. For instance, in the scenario proposed by \citet{DErcole2008}, a second generation forms in the central regions of a cluster. Furthermore, \citet{Vesperini2013} studied the long-term dynamical evolution of a multiple generations scenario with initial spatial segregation. Their $N$-body simulations indicate that the multiple populations do not fully mix spatially until a GC has lost $\sim60-70\%$ of its initial mass due to two-body relaxation \citep[see also][]{VHBetal2015, Miholics2015}. In this framework, the multiple populations of dynamically less evolved GCs, such as 47 Tuc, are not completely mixed and still exhibit an imprint of their initial distributions, whereas GCs that have lost a large fraction of their mass would be fully mixed spatially (e.g. NGC 6362 - \citet{Dalessandro2014}, \citet{Miholics2015}; M71 - \citet{Cordero2015a}).

It is however worth noting that the trend of a more centrally concentrated enriched population is maybe not universal. Using HST/WFC3 photometry, \citet{Larsen2015} found that the giants with normal composition are more centrally concentrated than the enriched giants in the inner parts of M15 (i.e. within the half-light radius). A similar reversed trend with respect to what was found in previous studies was also reported by \citet{Lim2016}, who found the CN-weak stars to be more centrally concentrated compared to the CN-strong stars in the GCs NGC 362 and NGC 6723, although their samples (based on ground-based narrow-band photometry) are incomplete in the central regions of the clusters. \citet{Larsen2015} showed that to explain their observed radial distributions in M15 in terms of dynamical mass segregation, giants with normal composition would need to be $\sim0.25~\msun$ more massive than their enriched counterparts. If this mass difference was due to a different initial helium composition for the different subpopulations, it would imply an extreme He enhancement for the enriched stars (Y~$\sim$~0.40), which is not compatible with the modest He enhancement deduced from the colour-magnitude diagram of M15 \citep[see][and discussions therein]{Larsen2015}. Whether the spatial distribution trend observed in the central regions of M15 then reflects initial conditions \citep{Larsen2015}, or to what extent other factors -  like a difference in the binary fraction of different chemical populations (see the paragraph below) or unaccounted observational systematic effects affecting the inferred radial distributions - could play a role however remains an open question \citep[see e.g.][]{VHB2015}. It will be interesting to compare future HST studies quantifying the spatial distribution of multiple populations in the inner parts of GCs with the behaviour reported in the above-mentioned studies.

Another piece of the puzzle comes from studies of binary star populations in GCs. Using $N$-body simulations to explore a scenario in which a second generation formed more centrally concentrated, \citet{Hong2015} predicted a higher binary disruption rate for ``second-generation" (Na-enriched) stars compared to ``first-generation" (Na-normal; ``field-like") stars. The first attempt to distinguish differences between binary fractions among multiple populations was done by \citet{D'Orazi2010} who found a lower binary fraction among Na-enriched stars by searching for Ba-enhanced giant stars in 15 GCs and radial velocity variations in NGC 6121. In absence of multi-epoch radial velocity measurements, Ba-enhanced stars allow to trace the incidence of binaries assuming the enrichment in Ba comes from s-process polluted material transfered from a former AGB binary companion. This assumption can be verified by modeling the s-process abundance pattern of Ba-enriched stars with AGB yields. For instance, \citet{Cordero2015b} confirmed from the analysis of many s-process elements that the abundance pattern of a Na-normal s-process enhanced star in 47 Tuc is consistent with enrichment by mass transfer from an 1.3 $\msun$ AGB companion. Clearly, s-process polluted stars that are/were part of a binary system can provide insights into multiple populations and binary evolution, but because these stars are extremely rare in GCs, their use for testing GC formation models is limited. A systematic and effective search for differences in binary fractions between Na-normal and Na-enriched stars was performed by \citet{Lucatello2015} on a sample of ten GCs. Their radial velocity campaign suggested that binary stars are more abundant among the Na-normal stars, which is consistent with the prediction by \citet{Hong2015} that the binaries of the Na-enriched population would be more efficiently destroyed if they were born in a denser environment (i.e. as part of a more centrally concentrated population). Given the lower mass at a given age of helium-enriched stars, binaries with enriched stars would also be easier to unbind (at the same encounter rate) than binaries with more massive ``helium-normal" stars, but the effect is expected to be small \citep{Lucatello2015}.

Besides chemical anomalies and intriguing spatial properties, the multiple populations of galactic GCs also display hints of kinematic differences in the outer regions of clusters (around the half-light radius and beyond). In a pioneering study of 20 GCs, \citet{Bellazzini2012} found a slightly lower line-of-sight velocity dispersion in the Na-enhanced groups of three GCs (NGC 6388, NGC 6441, NGC 2808) compared to stars with normal Na abundance. Furthermore, the Na-normal groups of three GCs in the \citet{Bellazzini2012} sample (NGC 2808, NGC 6171, and NGC 7078) seem to exhibit larger rotational amplitudes by a few \kms compared to Na-enhanced stars, although the differences reported were based on small samples and the statistical significance of these differences was not reported. Similarly, proper-motion and radial-velocity studies suggest that the enriched population of 47 Tuc may have a lower velocity dispersion \citep{Richer2013, Kucinskas2014}. Also in 47~Tuc, \citet{Richer2013} found that the presumably He-enriched stars (bluer main-sequence stars) have a more radially anisotropic velocity distribution (in the plane of the sky) based on HST proper motions in a field located around 2 half-light radii from the centre of the cluster. A similar signature was reported by \citet{Bellini2015} in NGC2808, also based on HST proper motions, where stars between $\sim1.5$ and 2 half-light radii and coinciding with the presumably He-enchanced populations show a more radially biased velocity distribution than the populations corresponding to normal non-enriched stars.

Connecting the kinematic and chemical properties of multiple populations in GCs offers a new approach for understanding how they formed. In particular, in the outer regions of clusters, where the relaxation timescale is longer and mixing is less important, we may expect to see imprints of the initial conditions and formation process of multiple populations. The simulations of \citet{Bekki2010, Bekki2011}, \citet{Mastrobuono2013, Mastrobuono2016} and \citet{VHBetal2015} have studied the synergy between the kinematic, spatial/morphological, and chemical properties of the multiple populations in GCs. In particular, \citet{VHBetal2015} explored what type of kinematic signatures would survive the long-term dynamical evolution of old GCs and may allow to distinguish different formation scenarios. Based on these results, it is worth emphasising that the kinematic differences reported so far between multiple populations in some GCs are not unique signatures of a specific formation scenario. The observational evidence collected in the outer parts of clusters is generally consistent with an enriched population forming more centrally concentrated, which is predicted in all the scenarios proposed to date. If two-body relaxation has not completely erased spatial and kinematic differences between subpopulations, the population that is more centrally concentrated is indeed expected to be dynamically cooler (i.e. lower velocity dispersion) and to have developed a more radially biased velocity distribution as it diffuses outwards. As pointed out by \citet{VHBetal2015}, a difference in the rotational amplitude of subpopulations is a potential signature that would provide further insight and possibly unique constraints to distinguish between formation scenarios, because the initial orbital configuration of normal and enriched stars implied by different scenarios can lead to opposite predictions for the sign of the difference in rotational amplitude.

In the present work, we study the GC M13 (NGC 6205) from a chemodynamical perspective and perform the first systematic search for kinematic differences between its multiple chemical populations. This old \citep[12.25 Gyr;][]{VandenBerg2013} and moderately metal-poor \citep[$\lbrack$Fe/H$\rbrack \sim -1.5$; e.g.][]{JohnsonPilachowski2012} GC exhibits relatively strong rotation \citep{Lupton1987, Cordero2014t, Fabricius2014}, has an Na-enriched and O-depleted population that is more centrally concentrated \citep{Lardo2011, JohnsonPilachowski2012}, as well as three chemical subpopulations \citep{JohnsonPilachowski2012}. The characteristics above, combined with the relatively large spectroscopic dataset of \citet{JohnsonPilachowski2012} from which precise radial velocities and uncertainties can be extracted, along with the hint of a difference in the rotational amplitude of its multiple populations \citep{Cordero2014t}, makes it an ideal target to perform such a study and introduce a robust statistical method to investigate the kinematics of multiple populations. We present the observations used in Section \ref{obs_data}, our kinematic analysis of subpopulations in Section \ref{kinematics}, a discussion of these results in Section \ref{discussion}, and summarise our findings in Section \ref{conclusions}.

\section{Observations}
\label{obs_data}

\subsection{Data reduction and analysis}

\begin{figure}
\centering
\includegraphics[width=9cm]{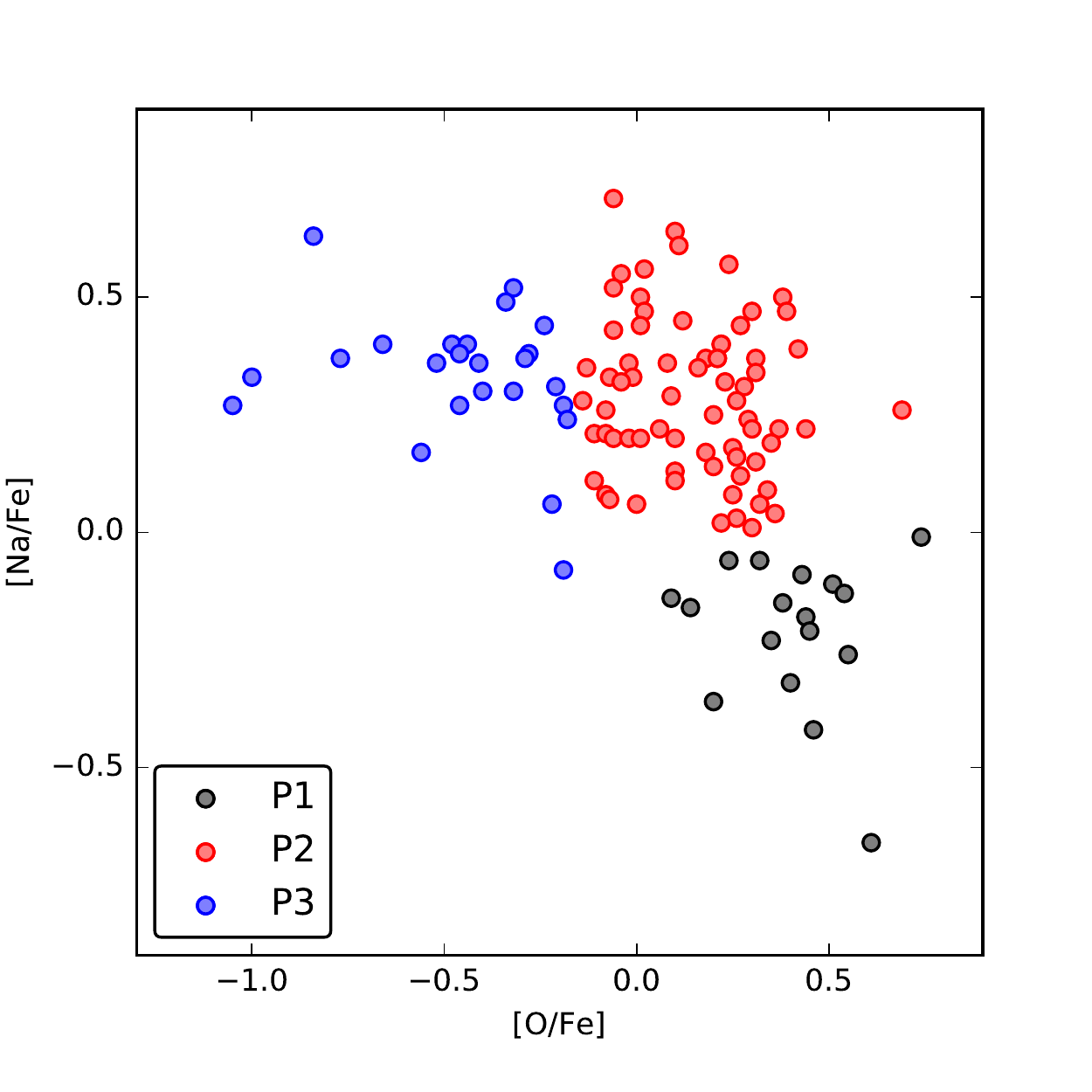}
\caption{Na-O anticorrelation for M13 giants using Na and O abundances from Johnson \& Pilachowski (2012). See Section \ref{chem_class} for the justification of the three chemical groups (P1, P2, and P3) identified.}
 \label{nao}
\end{figure}

{\renewcommand{\arraystretch}{1.6} 
\begin{table*}
\caption{Basic data and measured radial velocities for M13. This table is provided in its entirety in a machine-readable form in the online journal.}
\label{bestfit}
\begin{tabular}{l c c c c c c c c}
\hline
ID	& R.A. & Dec. & V &  $\langle RV \rangle$ & $\sigma_{\rm \langle RV \rangle}$ & [O/Fe] & [Na/Fe]$_{\mathrm{LTE}}$ & Population  \\
 & J2000 & J2000 &  & [km~s$^{-1}$] &  [km~s$^{-1}$] &  & & \\
\hline
CM	12	&	250.556994	&	36.554123	&	15.28	&$	-235.7	$&	0.5	&$	0.74	$&$ -0.01   $&         P1	\\
K	188	&	250.179091	&	36.461632	&	13.39	&$	-243.2	$&	0.4	&$	0.37 	$&$  0.22   $&	P2	\\
K	224	&	250.275644	&	36.422974	&	14.52	&$	-246.0	$&	0.4	&$	0.38 	$&$  0.50   $&	P2	\\
K	228	&	250.277032	&	36.47047	&	13.31	&$	-249.8	$&	0.4	&$	-0.06  $&$  0.52   $&	P2	\\
K	422	&	250.401823	&	36.285603	&	14.02	&$	-243.0	$&	0.4	&$	0.44 	$&$  0.22   $&	P2	\\
K	647	&	250.546518	&	36.306267	&	14.71	&$	-244.1	$&	0.5	&$	-0.07 	$&$  0.07   $&	P2	\\
K	656	&	250.56175	&	36.45554	&	13.04	&$	-241.7	$&	0.4	&$	0.26 	$&$  0.03   $&	P2	\\
K	659	&	250.568872	&	36.416309	&	14.24	&$	-240.3	$&	0.4	&$	-0.01 	$&$  0.33   $&	P2	\\
K	674	&	250.588623	&	36.564819	&	14.26	&$	-249.4	$&	0.4	&$	0.69 	$&$  0.26   $&	P2	\\
K	699	&	250.657823	&	36.403545	&	13.75	&$	-248.4	$&	0.4	&$	0.22 	$&$ 0.40    $&	P2	\\
L	6	&	250.291419	&	36.502876	&	15.00	&$	-251.5	$&	0.5	&$	-0.02 	$&$ 0.36    $&	P2	\\
L	16	&	250.312979	&	36.398304	&	14.60	&$	-245.6	$&	0.5	&$	0.44	$&$ -0.18   $&	P1	\\
L	18	&	250.313409	&	36.490002	&	13.78	&$	-249.8	$&	0.4	&$	0.10 	$&$  0.11   $&	P2	\\
L	26	&	250.32086	&	36.429989	&	13.58	&$	-242.5	$&	0.4	&$	-0.14 	$&$ 0.28    $&	P2	\\
L        29	&        250.323850	&        36.491638	&        14.51 	&$    	-238.1           $&      0.5     &$     ...        $&$ 0.52    $&        P2/P3  \\
L	32	&	250.327637	&	36.478725	&	15.12	&$	-242.3	$&	0.5	&$	0.01 	$&$ 0.50    $&	P2	\\
\hline
\end{tabular}
\end{table*}
}

The spectra were acquired using the WIYN-Hydra fiber positioner (R $\approx$ 18,000) on 2011 May 19-20 \citep{JohnsonPilachowski2012} and have wavelength coverage ranging from 6050–6350 \AA . The data reduction process \citep[see also][]{JohnsonPilachowski2012} was accomplished with standard IRAF tasks, which included using the ccdproc routine to trim and bias subtract the raw frames.  The dohydra routine was used to identify and trace the fibers, apply the flat-field correction, measure and apply the wavelength calibration, remove cosmic rays, extract individual fibers, and subtract the sky spectra. Several exposures for each star (ranging from 4 to 12 in our program) were taken to build high S/N spectra, which have been previously used to determine Fe, Na, and O abundances by \citet{JohnsonPilachowski2012}. Individual exposures had signal-to-noise ratios (S/N) ranging from about 25-50.

Relative radial velocities were measured on single exposures by cross-correlating the object spectrum with a smoothed synthetic spectrum of similar atmospheric parameters. The synthetic spectra used as a template are interpolated models from the grids of \citet{Coelho2005}. We used the cross-correlation task {\it fxcor} in IRAF to determine a cross-correlation profile, whose strongest peak represents the radial velocity of the star. {\it Fxcor} allows to interactively select wavelength regions that are clear of bad pixels and telluric features. For each star the strongest peak of the cross-correlation function was well fit by a Gaussian. A quantitative estimate of the quality of the Gaussian fit, based on the Tonry-Davis R-value \citep{TonryDavis1979}, is given by {\it fxcor} in addition to the velocity measurement and its uncertainty. The R-value corresponds to the ratio of the strongest cross-correlation peak height to the average height of unselected peaks. As a rule of thumb, velocities with an R-value $\geq$ 5 are regarded as reliable. In our sample the R-values were always greater than 12 (and typically 30), ensuring that the velocities included in our analysis are reliable. Heliocentric corrections were applied to the relative velocities obtained for each exposure using the IRAF task {\it rvcorrect}. Radial velocities determined from different exposures of the same star were found to be in agreement. Finally, an error-weighted average heliocentric velocity with its error was computed for each star. The weighted average heliocentric velocity of the whole sample is $-$244.8 km s$^{-1}$ (standard deviation 6.1 km s$^{-1}$), which is in agreement with the systemic velocity of $-$246.2 km s$^{-1}$ reported by \citet{Lupton1987} and with  the mean heliocentric radial velocity of $-$244.2 km s$^{-1}$ (standard deviation 7.1 km s$^{-1}$) taken from the catalogue of \citet[][2010 edition]{Harris1996}. These are also in agreement with our Bayesian determination of the mean velocity of the full sample (see Section \ref{kinematics}).

Table 1 presents the identifications from \citet{Ludendorf1905} and \citet{Kadla1966}, coordinates, V magnitudes, error-weighted average heliocentric radial velocities, radial velocity errors, chemical abundance ratios for Na and O from \citet{JohnsonPilachowski2012}, and chemical population classification (see next section) for a sample of 113 giant stars.

\subsection{Chemical classification}
\label{chem_class}

All the stars in our sample have proper motion membership probabilities higher than 70\%\footnote{Proper motion membership data was obtained from \citet{Cudworth1979}.}, radial velocities within 3$\sigma$ from the cluster's mean radial velocity, and uniform [Fe/H] ratios within the errors, ensuring membership to the cluster. To assess the possible field contamination rate in our sample, we used the Besan\c{c}on Galaxy model \citep{Robin2003} to create a synthetic CMD and velocity distribution along lines of sight near M13.  This exercise returned a 0.05$\%$ probability that a random field star would have a velocity, metallicity, surface gravity, and color that is similar to those expected for M13 giants.  Combined with our previous membership vetting using the Cudworth \& Monet (1979) proper motions, the Besan\c{c}on simulation suggests that field star contamination in our sample is negligible.
M13 has an extended Na-O anticorrelation as can be seen in Figure~\ref{nao}. To distinguish its multiple populations, we separated the Na-normal population, i.e. stars with halo-like chemical composition ([Na/Fe]$<0$ and [O/Fe]$>0$), from the Na-enriched stars. We refer to the Na-normal population as P1. The Na-enriched stars constitute the majority of the sample and span a wide range of [O/Fe] abundance ratios ($\sim$1.5 dex). The Na-enriched stars follow a continuous distribution in the Na-O plane, which is likely a consequence of the spectroscopic errors (typically 0.15 dex in [Na/Fe] and [O/Fe] for the stars in our sample), challenging the identification of different chemical groups. To overcome this issue we used a k-means clustering algorithm on the Na-enriched stars to separate the O-depleted stars from the extremely O-depleted stars. The k-means algorithm is a simple unsupervised machine-learning algorithm that randomly assigns cluster centers to the data; the cluster centers are refined by determining the vector mean of the initial clusters in an iterative process until convergence, allowing the classification of the full data set. The cluster membership of each data point is determined by minimizing the distance to the final cluster centers. These two groups were named P2 (Na-enriched and O-depleted stars) and P3  (Na-enriched and extremely O-depleted stars). Notice that two stars in our sample have only Na or O abundance measurements, but not both. In particular star L1050 does not have a Na estimate; however, its O abundance allows to classify this star as either P1 or P2 population. Similarly, star L 29 does not have an O abundance, thus, L 29 can be classified as either P2 or P3.

\begin{figure*}
\centering
\includegraphics[width=18cm]{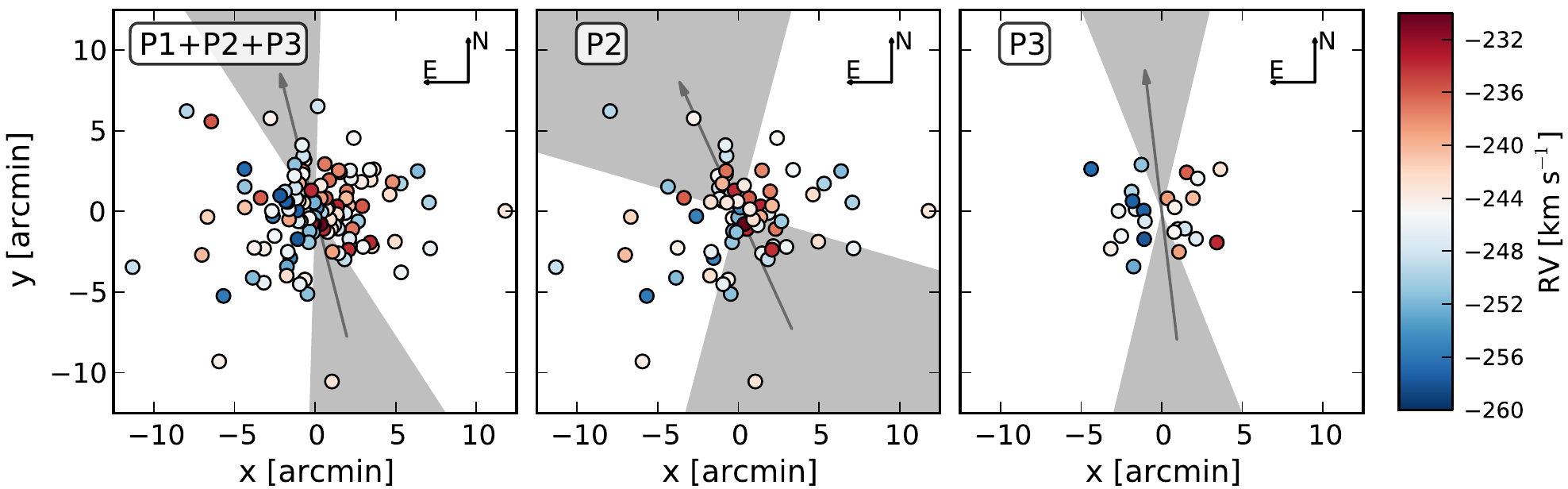}
\caption{Radial velocity map of the different subsamples (from left to right: full sample P1+P2+P3, ``intermediate" P2, ``extreme" P3) showing the position of the stars in projected cartesian coordinates and the position angle of the inferred rotation axis (grey arrow) with 1-$\sigma$ confidence intervals (shaded grey region). The radial velocity of each star is indicated by its color, as shown in the color bar. We do not show the velocity map of the ``normal" (P1) subsample because it has a small number of stars and its putative rotation axis is poorly constrained.}
 \label{vel_map}
\end{figure*}

\section{Kinematic analysis}

\label{kinematics}

\subsection{Method}

We describe in this section a Bayesian method aimed at quantifying the kinematic properties of the multiple chemical populations in the GC M13. We use a discrete fitting technique to compare a simple kinematic model (including a radial depdendence of the rotational amplitude and velocity dispersion of the cluster) to individual radial velocities. We apply this method to the chemical subsamples P1, P2, and P3 defined in the previous section, as well as to the sample as a whole (P1+P2+P3) and to a combination of subsamples P1 and P2. Radial velocity maps of the three main samples discussed below (P2, P3, and P1+P2+P3) are shown in Figure~\ref{vel_map}. An indication of rotation can already be seen just by looking at these velocity maps and noticing the velocity gradient across the cluster.

According to Bayes theorem, the posterior probability distribution of the model parameters $\Theta$, given the data $\xvec$, is given by

\begin{equation}
p(\Theta|\xvec) = \frac{p(\xvec|\Theta) p(\Theta) }{p(\xvec)} \ ,
\end{equation}
where $p(\Theta)$ is the prior probability distribution, $p(\xvec|\Theta)$ is the likelihood function, and $p(\xvec)$ is the evidence. In the context of the present analysis, the latter is a normalisation factor which we will simply neglect. In the cases considered here, the data $\xvec$ is a set of $N$ stars with radial velocities $v_i$ (and uncertainties $\sigma_{v, i}$) and known on-sky coordinates.

The probability of the data given the model parameters, i.e. the likelihood function, is expressed as
\begin{equation}
p(\xvec|\Theta) = \prod_{i=1}^{N} \Lambda_i(\xvec_i|\Theta) \ ,
\label{Eq_Lambda}
\end{equation}
which is the product of individual likelihood functions $\Lambda_i(\xvec_i|\Theta)$ calculated for every star in the sample. The likelihood function for the radial velocity of individual stars depends on our assumption for the rotation curve of the cluster as well as its velocity dispersion profile. For the velocity dispersion profile, we assume the functional form of the \citet{Plummer1911} model, defined by its central velocity dispersion $\sigma_0$ and its scale radius $a$:

\begin{equation}\label{sigma2R}
\sigma^2(R) = \frac{\sigma_0^2}{\sqrt{1 + R^2/a^2}} \ ,
\end{equation}
where $R$ is the projected distance from the centre of the cluster. For the family of projected \citet{Plummer1911} models, $a$ is equal to the half-mass radius, which is equivalent to the half-light radius if the mass-to-light ratio is constant throughout the cluster. Note that this is not necessarily the case for GCs, which are typically mass segregated due to their collisional nature, as appears to be the case for M13 \citep[e.g.][]{Goldsbury2013}. We stress that our goal here is not to perform a detailed and self-consistent dynamical analysis of M13, but to search for possible differences in the kinematics of chemical subpopulations. In this purely kinematic approach, we thus chose the Plummer model merely for its convenient, simple, yet flexible parametrisation and analytic description of a system with a constant-density core. As we illustrate below, this parametrisation provides a satisfying description of the observed velocity dispersion profile of the cluster (and of the different chemical subsamples considered) given the size of our dataset and the quality of the data (see Section~\ref{results}).

{\renewcommand{\arraystretch}{1.6} 
\begin{table*}
\caption{Median and $\pm1 \, \sigma$ uncertainties for the six free parameters of the kinematic analysis of different chemical subsamples.}
\label{bestfit}
\begin{tabular}{l c c c c c c c c}
\hline
& Sample & $N_ {\rm stars}$ & $v_{0}$ &  $A_{\rm rot}$ & $R_{\rm peak}$ & PA$_{0}$ & $\sigma_0$ & $a$  \\
& & & [km~s$^{-1}$] &  [km~s$^{-1}$] & [arcmin] & [$^{\circ}$] & [km~s$^{-1}$] & [arcmin]  \\
\hline
P1 & ``normal" halo-like population & 17 &$-243.9^{+1.8}_{-1.9}$  & $2.4^{+2.7}_{-1.7}$ & $2.9^{+2.2}_{-1.9}$ & $56^{+53}_{-134}$ & $8.8^{+2.5}_{-1.9}$ & $3.0^{+1.9}_{-1.6}$ \\
P2 & ``intermediate" population& 70 & $-245.0^{+0.7}_{-0.7}$ & $1.3^{+1.1}_{-0.9}$ & $2.5^{+2.2}_{-1.4}$ & $24^{+49}_{-39}$ & $7.0^{+1.4}_{-0.9}$  & $3.1^{+1.8}_{-1.4}$ \\
P3 & ``extreme" population& 24 &  $-246.1^{+1.2}_{-1.1}$ &  $5.6^{+1.7}_{-1.6}$ &  $3.0^{+2.1}_{-1.7}$ &  $7^{+15}_{-20}$ &  $6.1^{+1.6}_{-1.1}$ &  $3.7^{+1.6}_{-1.9}$ \\
P1+P2 & ``normal" + ``intermediate"& 88 & $-244.7^{+0.6}_{-0.6}$ & $1.4^{+1.0}_{-0.9}$&$2.1^{+2.3}_{-1.2}$ &$33^{+57}_{-37}$ &$7.0^{+1.4}_{-0.8}$ & $3.3^{+1.7}_{-1.5}$ \\
P1+P2+P3 & full sample& 113 & $-244.9^{+0.5}_{-0.5}$ & $2.7^{+0.9}_{-0.8}$ & $ 1.5^{+1.0}_{-0.6}$  &$14^{+19}_{-16}$  & $6.6^{+0.8}_{-0.6}$  & $4.1^{+1.3}_{-1.4}$  \\

\hline
\end{tabular}
\end{table*}
}

\begin{figure*}
\centering
\includegraphics[width=17cm]{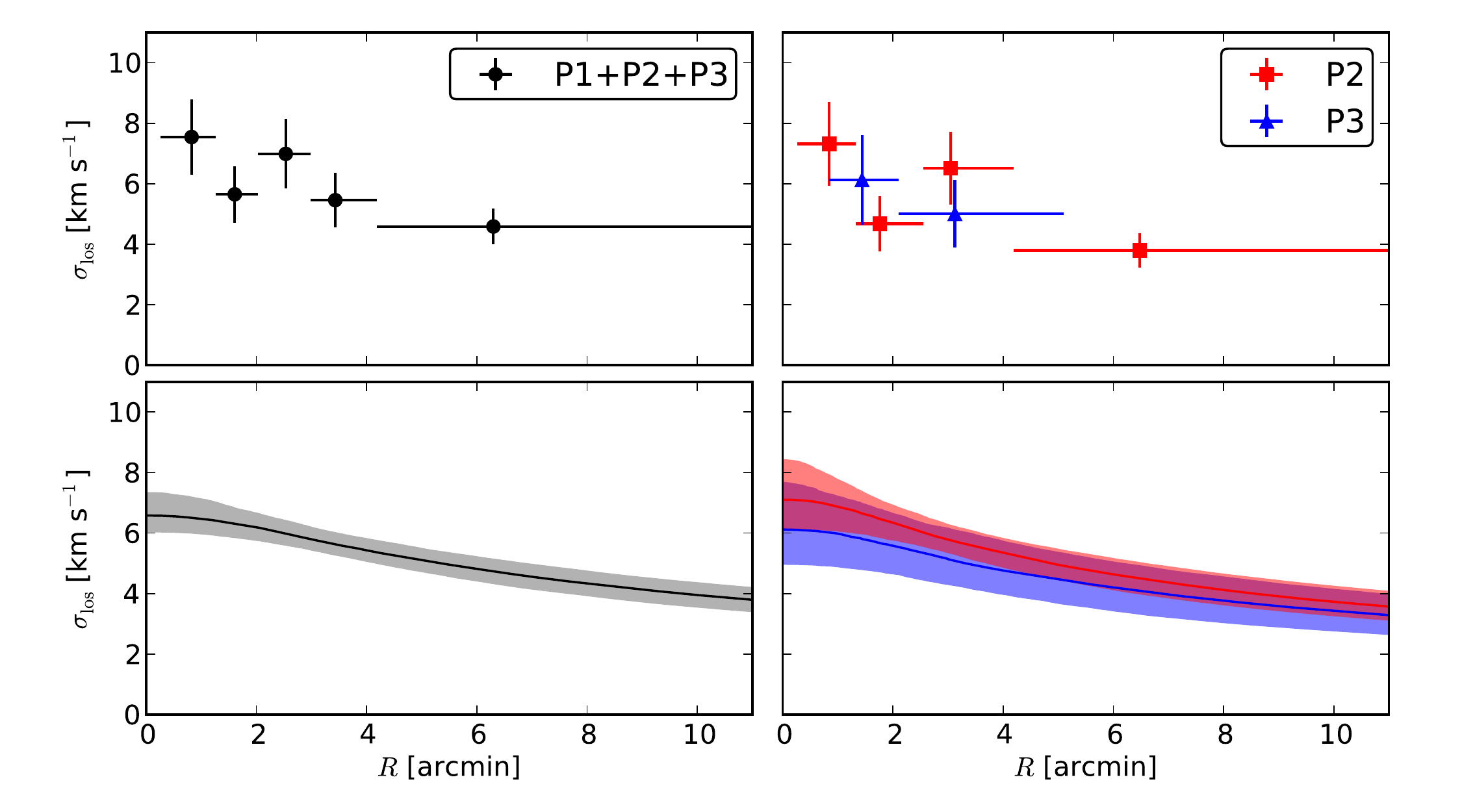}
\caption{Line-of-sight velocity dispersion profile as a function of projected distance from the centre of the cluster for the subsamples P1+P2+P3 (left panels), P2 and P3 (right panels). The upper panels show the profiles obtained from binning the data, and the lower panels show the best fits as solid lines with shaded regions illustrating the 1-$\sigma$ uncertainty envelopes of the dispersion profiles. Note that the fits were performed on the discrete velocities, and the binned profiles are shown here only for illustration purposes.}
\label{sigma_profiles_P2_P3}
\end{figure*}

\begin{figure}
\centering
\includegraphics[width=8.5cm]{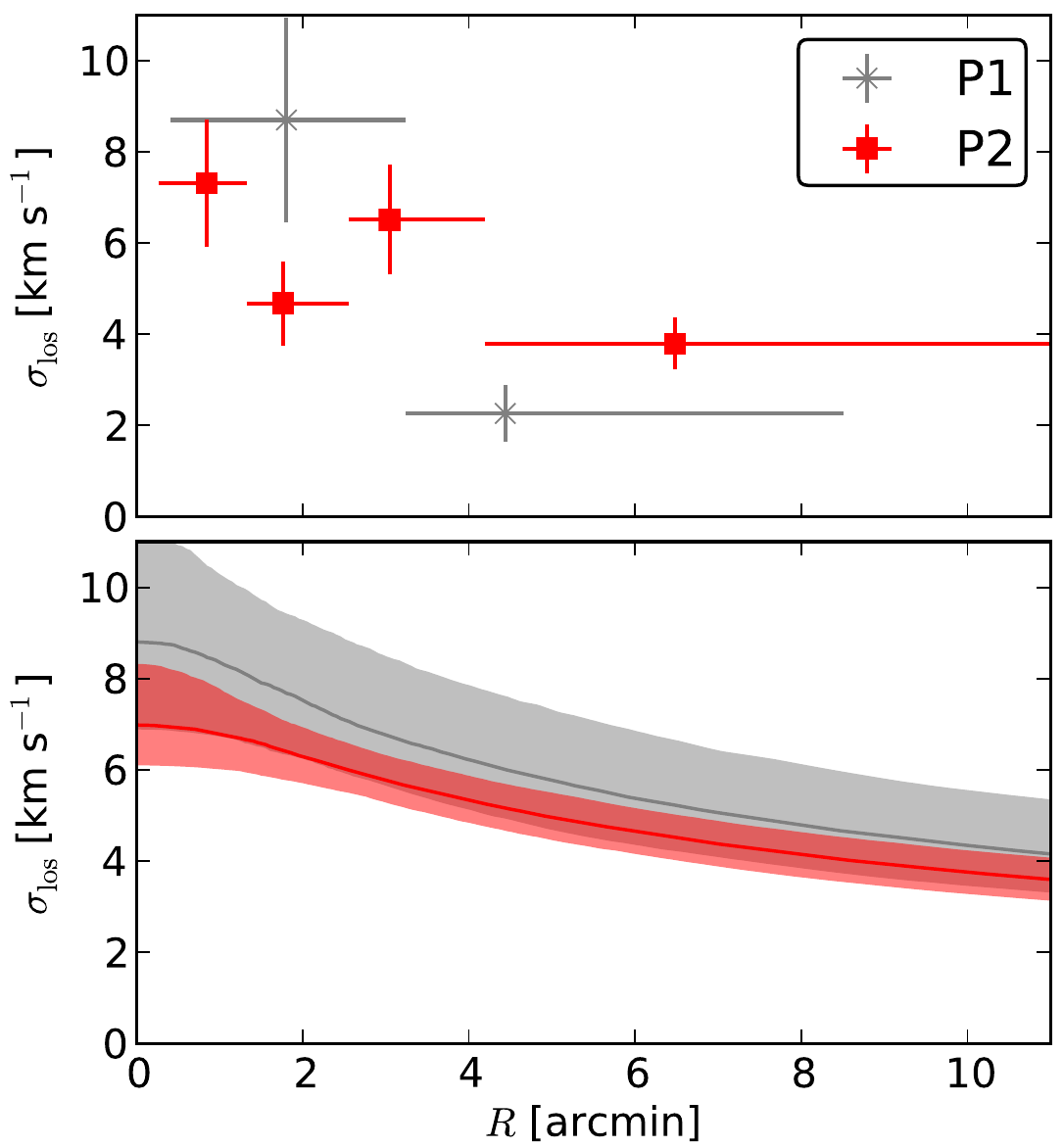}
\caption{Same as the right panels of Figure \ref{sigma_profiles_P2_P3}, but comparing the line-of-sight velocity dispersion profiles of subsamples P1 and P2.}
\label{sigma_profiles_P1_P2}
\end{figure}

\begin{figure*}
\centering
\includegraphics[width=17cm]{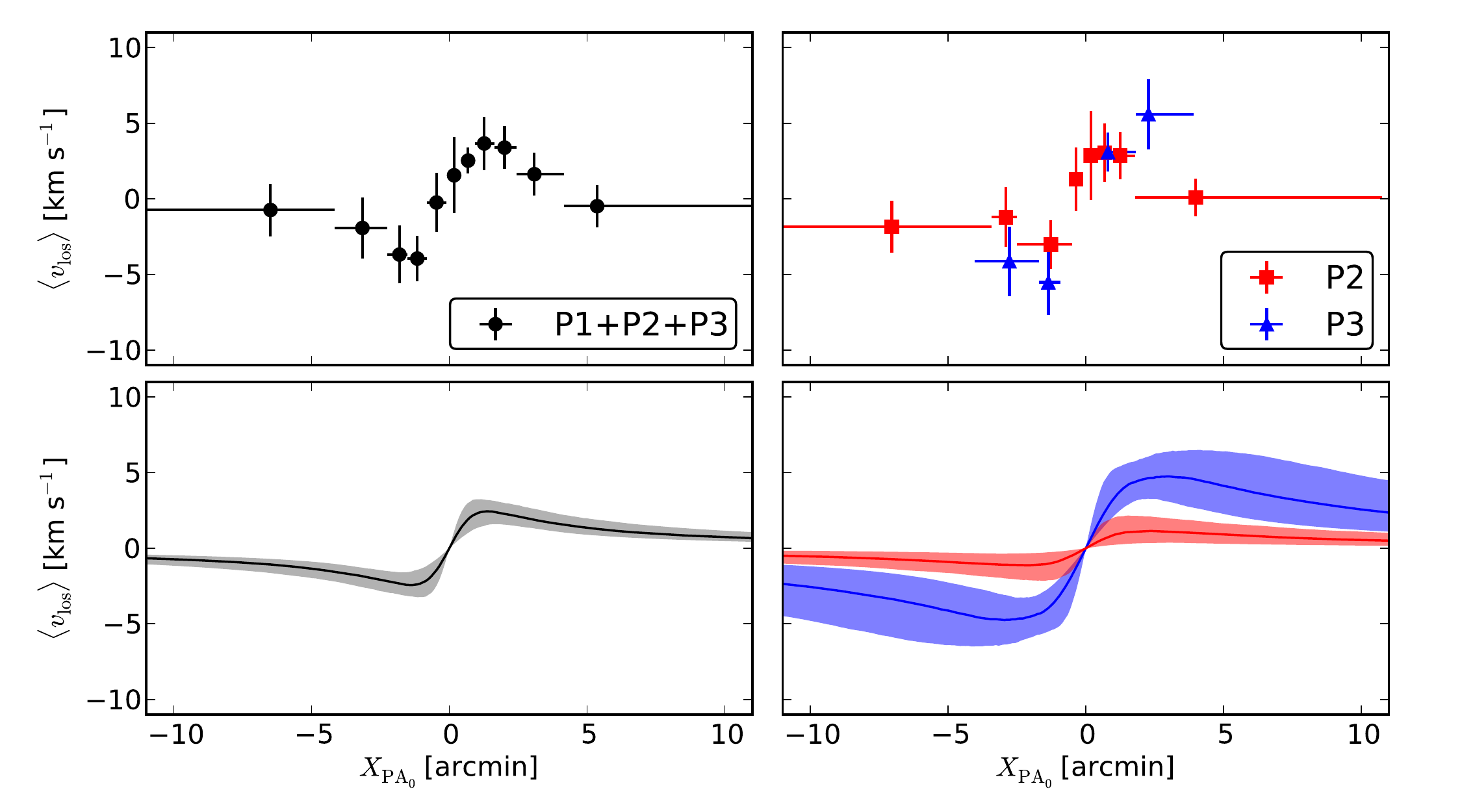}
\caption{Mean line-of-sight velocity as a function of projected distance from the rotation axis (i.e. rotation curve) for the subsamples P1+P2+P3 (left panels), P2 and P3 (right panels). The upper panels show the rotation curves obtained from binning the data, and the lower panels show the best fits as solid lines with shaded regions illustrating the 1-$\sigma$ uncertainty envelopes of the rotation curves. Note that the fits were performed on the discrete velocities, and the binned profiles are shown here only for illustration purposes.}
\label{rot_curves_P2_P3}
\end{figure*}

\begin{figure}
\centering
\includegraphics[width=8.5cm]{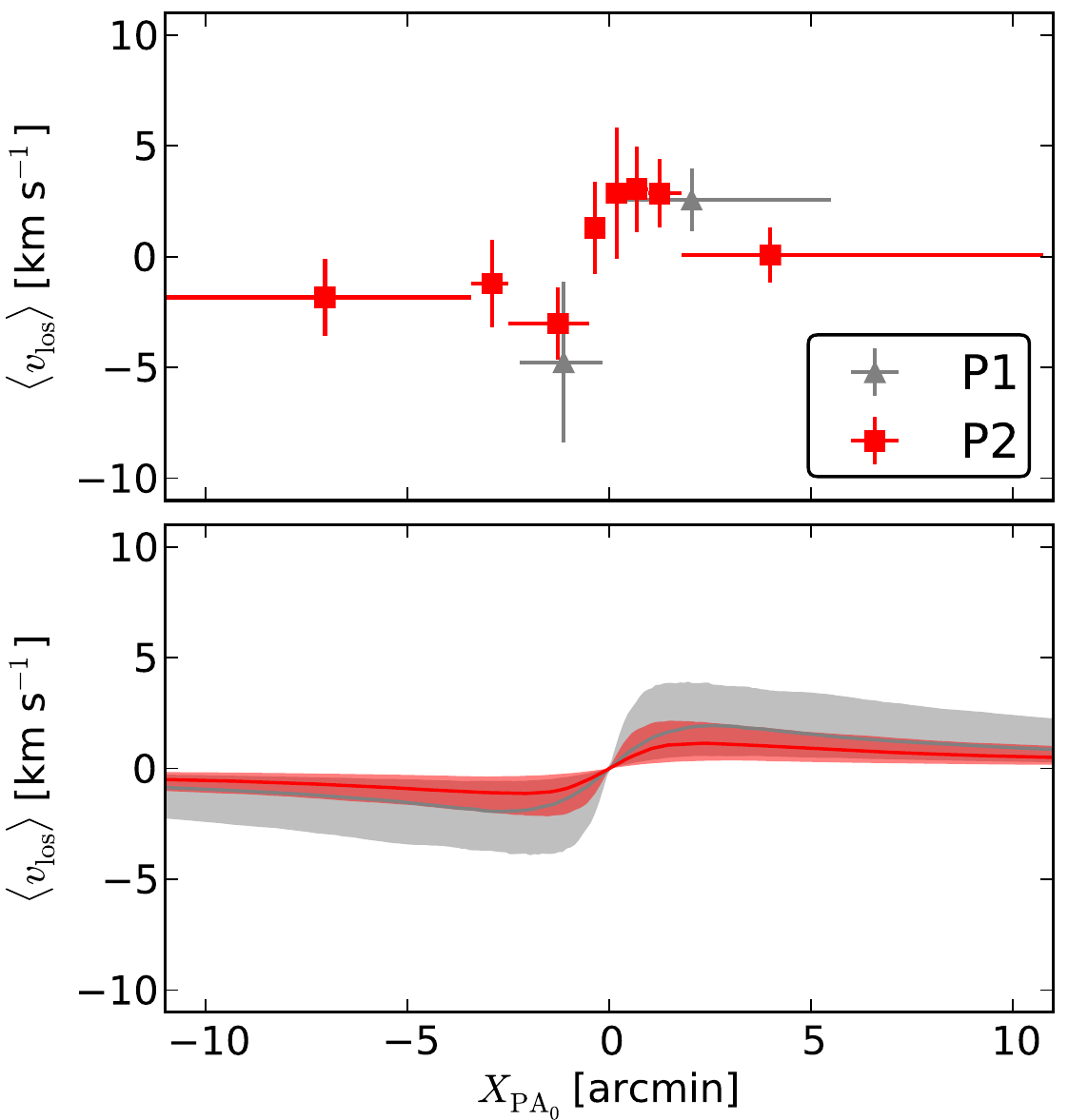}
\caption{Same as the right panels of Figure \ref{rot_curves_P2_P3}, but comparing the rotation curves of subsamples P1 and P2.}
\label{rot_curves_P1_P2}
\end{figure}

\begin{figure}
\centering
\includegraphics[width=9cm]{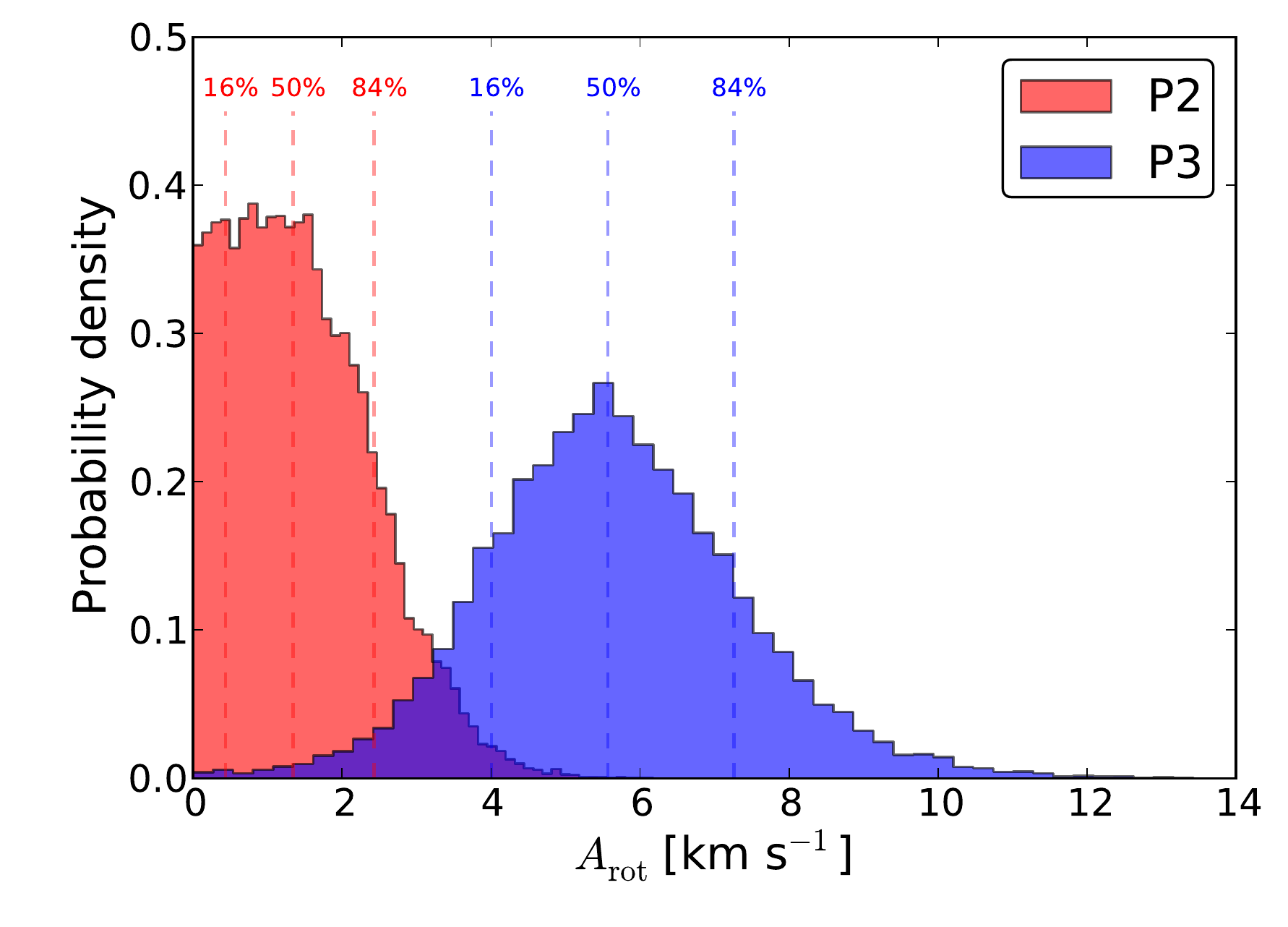}
\caption{Posterior probability distribution of the rotational amplitude $A_{\rm rot}$ for the subsamples P2 and P3. The vertical dashed lines indicate 16\%, 50\%, and 84\% percentiles for each probability distribution.}
 \label{post_Arot_P2_P3}
\end{figure}

\begin{figure}
\centering
\includegraphics[width=9cm]{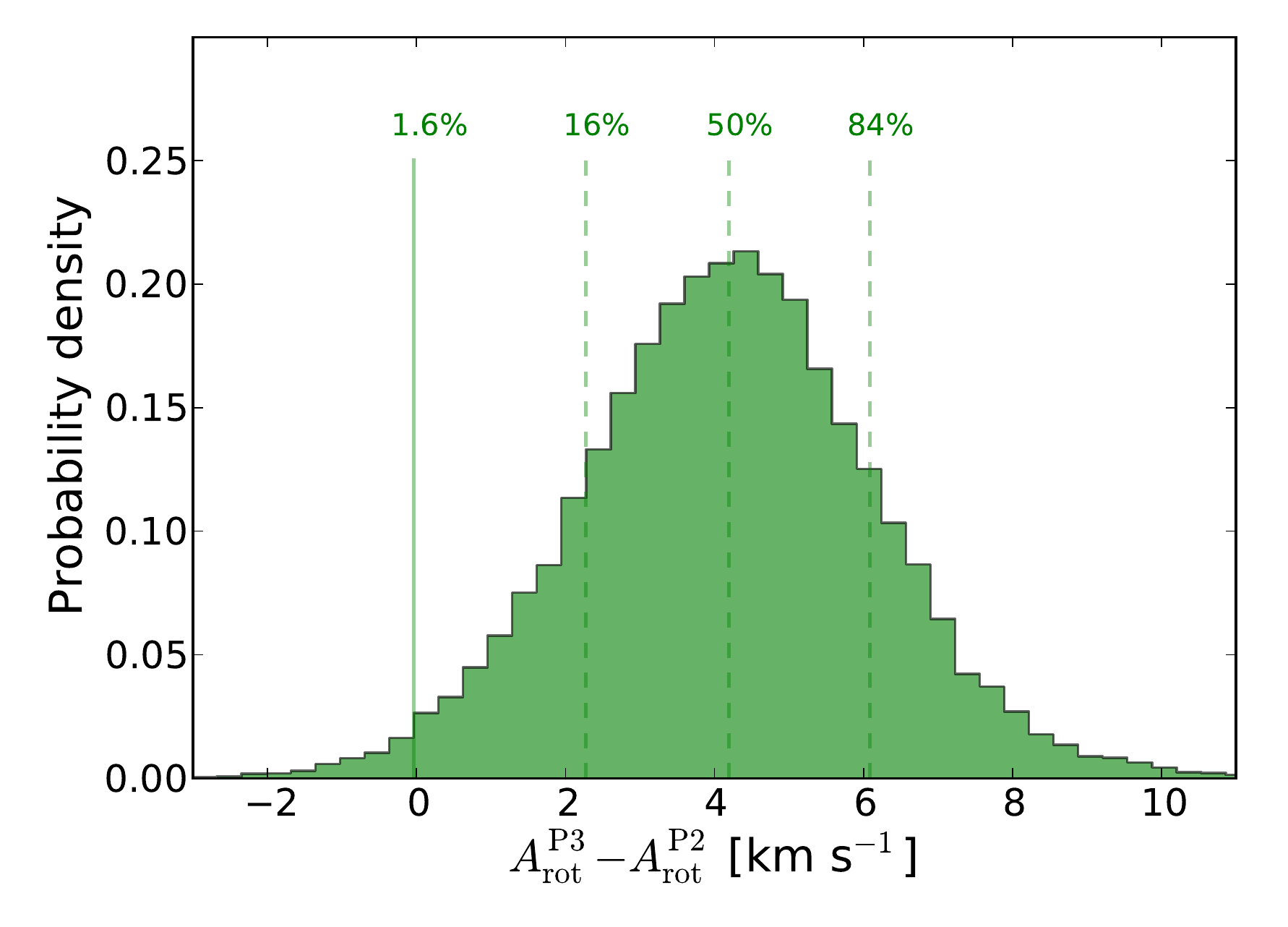}
\caption{Posterior probability distribution of the difference in rotational amplitude between the subsamples P3 and P2. The vertical dashed lines indicate 16\%, 50\%, and 84\% percentiles of the probability distribution. The solid vertical line indicates that the probability that the difference is larger than 0 \kms \ is 98.4\%.}
 \label{post_P2_P3}
\end{figure}

\begin{figure*}
\centering
\includegraphics[width=17cm]{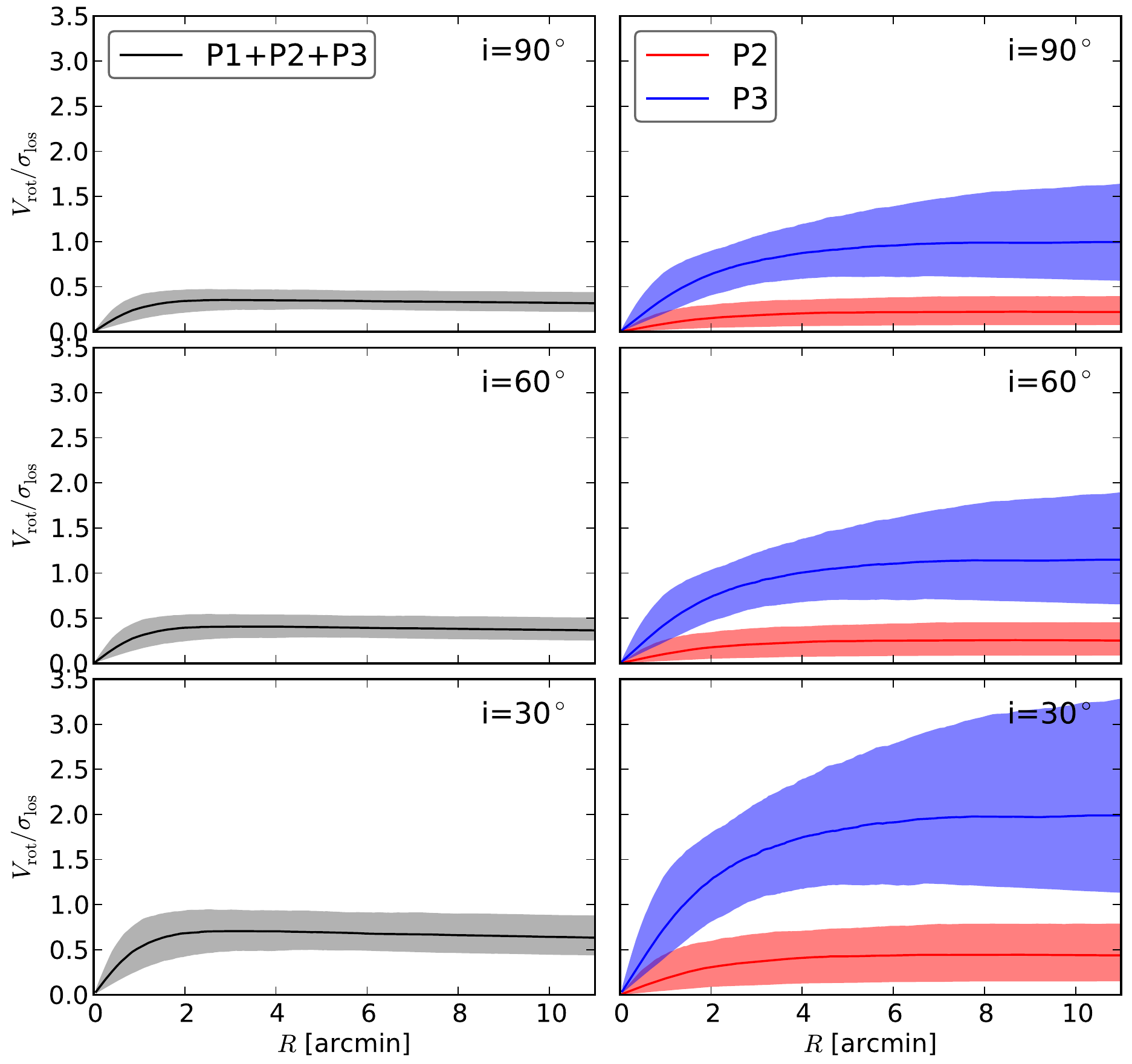}
\caption{Ratio of the mean rotational amplitude over the one-dimensional velocity dispersion as a function of projected distance from the centre of the cluster for the subsamples P1+P2+P3 (left panels), P2 and P3 (right panels), and adopting an inclination angle of 90$^{\circ}$, 60$^{\circ}$, or 30$^{\circ}$ (from top to bottom) between the rotation axis and the line of sight. Best fits are shown as solid lines with shaded regions illustrating the 1-$\sigma$ uncertainty envelopes of the $V_{\rm rot}/\sigma_{\rm los}$ profiles.} 
\label{vrot_sigma}
\end{figure*}

For the rotation curve, we assume cylindrical rotation and adopt a functional form inspired by the outcome of violent relaxation \citep{Gott1973, LB1967}, and parametrized in the following way \citep[e.g.][]{HB2012, Mackey2013}:

\begin{equation}
V_{\rm rot}  \sin{i}(X_{\rm PA_{0}}) = \frac{2 \ A_{\rm rot}}{R_{\rm peak}} \ \frac{X_{\rm PA_{0}}}{1 + (X_{\rm PA_{0}}/R_{\rm peak})^2} \ ,
\end{equation}
where $V_{\rm rot}  \sin{i}$ is the rotational amplitude along the line of sight at a projected distance $X_{\rm PA_{0}}$ from the rotation axis, and $A_{\rm rot}$ is the peak rotational amplitude (again along the line of sight) occurring at projected distance $R_{\rm peak}$ from the centre. We define the position angle (with respect to the centre of the cluster) as increasing anti-clockwise in the plane of the sky from North (${\rm PA} = 0^{\circ}$) towards East (${\rm PA} = 90^{\circ}$). We adopt negative values of $V_{\rm rot}  \sin{i}$ for position angles between PA$_{0}$ and PA$_{0}+180^{\circ}$, where PA$_{0}$ is the position angle of the rotation axis in the plane of the sky.  Given that the inclination of the rotation axis with respect to the line of sight is unknown, $V_{\rm rot} \sin{i}$ is a lower limit to the actual three-dimensional rotational amplitude ($V_{\rm rot}$). If the rotational axis was aligned with the line of sight, rotation would be in the plane of the sky and we would obviously miss any rotational signature when using radial velocities. With solid-body rotation in the inner parts and a rotation curve peaking at intermediate radii (around the half-mass radius) followed by a decline, the functional form above conveniently captures the general behaviour seen in self-consistent equilibrium models of rotating stellar systems \citep[e.g.][]{Lagoute1996, VarriBertin2012} as well as in evolutionary models of rotating globular clusters \citep[e.g.][]{Kim2002, Fiestas2006, Hong2013}. As we show below, this parametrisation also satisfyingly reproduces the observed rotation curve of the cluster (as well as the rotation curve of the different chemical subsamples considered) obtained from binning the radial velocity data (see Section~\ref{results}).

%Flattening is small, so we do not correct for flattening along the minor axis when fitting the rotational signal or building the rotation curve from binning the radial velocity data.

To avoid projection effects due to the non-negligible angular diameter and distance from the equatorial plane of M13, we project the coordinates of each star on the plane of the sky along the line-of-sight vector through the cluster centre \citep[e.g.][]{VdV2006}, and we use those projected cartesian coordinates for our kinematic analysis.\footnote{To speed up the exploration of parameter space, the projected distance $X_{\mathrm{PA}_0}$ to the axis of rotation for each star can then be parametrized as its $x_i'$ component in a rotated frame $\mathbf{x}_i'=\mathbf{R}(\mathrm{PA}_0)\mathbf{x}_i$,
where $\mathbf{R}(\mathrm{PA}_0)$ is a rotation matrix in two dimensions.}

Assuming a Gaussian distribution of velocities, the likelihood function for an individual star is then given by 

\begin{equation}
 \Lambda_i(\xvec_i|\Theta) = \frac{1}{\sqrt{2 \ \pi  \ (\sigma_{v, i}^2 + \sigma^2)}} \exp{\left[\frac{ - ( v_i - v_{0} - V_{\rm rot}  \sin{i})^2}{2 \ (\sigma_{v, i}^2 + \sigma^2)}\right]}  \ ,
\end{equation}
where $v_0$ is the systemic radial velocity of the cluster, with the other variables defined as above. Because our likelihood function is based on the conditional probability of a radial velocity measurement given the position of a star, our fitting procedure is not biased by the spatial sampling of the stars in our sample and subsamples: the kinematics is simply better constrained in regions of the cluster that are better sampled, and less well constrained in regions that are poorly sampled. Note that we do not include a contamination term in our likelihood function since potential non-members have already been excluded from our dataset.

Given that some metal-poor red giants within $\sim0.5-1.0$~mag from the red giant branch tip are known to exhibit a velocity jitter at the level of $\sim 1.5-2.0 \ {\rm km~s}^{-1}$ \citep{Carney2003}, we also consider an additional broadening term in the Gaussian likelihood function above, such that $\sigma_{v, i}^2 \rightarrow \sigma_{v, i}^2 + (2 \ {\rm km~s}^{-1})^2$ for bright giants within 1 mag from the red giant branch tip of M13 (V=12). This however only affects a fraction of our sample, and we performed additional experiments by running the same analysis but without including this velocity jitter term, which revealed that neglecting this effect would not affect our results and conclusions. 

The model parameters that we vary to find the best match to the data are $\Theta = \{ v_0, A_{\rm rot}, R_{\rm peak}, {\rm PA}_0, \sigma_0, a\}$. For those parameters, we adopt uninformative uniform priors over the following ranges: $-280 < v_0 < -220 \ {\rm km~s}^{-1}, 0 < A_{\rm rot} < 14 \ {\rm km~s}^{-1}, 0.5 < R_{\rm peak} < 6 \ {\rm arcmin},  -180 < {\rm PA}_0 < 180^{\circ}, 0 < \sigma_0 < 14 \ {\rm km~s}^{-1}, 0.5 < a < 6 \ {\rm arcmin}$. The boundaries were chosen to bracket a wide range of possible values for our dataset. In particular, we expect $R_{\rm peak}$ and $a$ to be comparable (within a factor of a few) to the half-light radius of M13 (1.69 arcmin based on the 2010 version of the Harris catalogue; \citet{Harris1996}). Throughout the analysis, we keep the position of the centre of M13 fixed at $\alpha$ = 16$^{\rm h}$41$^{\rm m}$41.634$^{\rm s}$, $\delta$ = $+$36$\degr$27'40.75'' (J2000).

To explore the parameter space and efficiently sample the posterior probability distribution for the parameters above, we use {\tt emcee} \citep{FM2013}, a {\sc python} implementation of the affine invariant Markov chain Monte Carlo (MCMC) ensemble sampler of \citet{GW2010}. We typically consider 500 walkers, and for each of these 800 steps in parameter space. Convergence is generally obtained within a few hundred steps or less, and we consider a burn-in phase of 600 steps for all our fits to be on the safe side. We also checked that our results are robust against changes to the initial positions of the walkers.

The method described above presents clear advantages over the traditional method used for detecting/characterising rotation in GCs \citep{Cote1995, Bellazzini2012}. The latter consists of dividing the line-of-sight velocity dataset in two halves by a line passing through the centre with a given position angle, and computing the mean line-of-sight velocity for each subsample. The position angle of the dividing line is varied and the difference between the mean velocities is plotted against position angle, and then the pattern is fitted with a sine function. The position angle for which the difference in mean velocities is maximised gives an estimate of the position angle of the rotation axis, and the amplitude of the sine curve gives an indication of the mean rotational amplitude. It is however difficult to obtain reliable uncertainties from that method, because the data points shaping this sine pattern are all correlated (the mean velocity difference for each data point is computed from the radial velocities of the whole sample). Our discrete method deals in a straightforward and reliable way with the uncertainties on all fitting parameters, it avoids loss of information from binning the data, and it has the advantage of treating the velocity dispersion and rotational components simultaneously when comparing kinematic models to the data.

\subsection{Results}
\label{results}

Table \ref{bestfit} lists, for the different subsamples considered and for the six free parameters of our analysis, the median value from the posterior probability distribution function (marginalised over all other parameters), with error bars indicating the interval enclosing the central 68\% of the probability distribution. We also show in Figures \ref{post_all}, \ref{post_intermediate}, \ref{post_extreme}, \ref{post_primordial}, and \ref{post_intprim} of Appendix \ref{appendixA} the two-dimensional projections of the posterior probability distribution on the planes determined by every pair of parameters. These figures also show histograms of the marginalised posterior probability distribution for each parameter of the kinematic model.

Figures \ref{sigma_profiles_P2_P3} and \ref{sigma_profiles_P1_P2} show a comparison of the line-of-sight velocity dispersion profiles of different subsamples (P1+P2+P3, P1, P2, P3). The binned dispersion profiles are obtained by assuming an equal number of stars in each bin (apart from the outermost bin in which any additional leftover stars are also included) and using the Maximum Likelihood estimator of \citet{PM1993} to compute the dispersion in each bin and its uncertainty. For the radius of each bin, we adopt the mean radius of all the stars in that bin, and for the error bar on the radius, we simply assume that they extend over the whole radius range spanned by the stars in a given bin. Recall that the fits were performed on the discrete radial velocity measurements, so these binned profiles are shown here only for illustration purposes. Best-fit dispersion profiles inferred from our fitting procedure are shown in the bottom panels of Figures \ref{sigma_profiles_P2_P3} and \ref{sigma_profiles_P1_P2}, with shaded regions showing 1-$\sigma$ envelopes. To obtain these 1-$\sigma$ envelopes, we first sampled 1000 sets of parameters from the post burn-in phase of a given MCMC chain. From those, we then compute the associated line-of-sight velocity dispersion profile. Then, at a given radius, we find the 16\% and 84\% percentiles of the dispersion of all 1000 dispersion profiles, and repeat for different radii. The solid lines representing the best-fit profiles represent the median of all the computed profiles at a each radius.

Similarly, Figures \ref{rot_curves_P2_P3} and \ref{rot_curves_P1_P2} show a comparison of the rotation curves (mean line-of-sight velocity as a function of projected distance from the rotation axis) of different subsamples (P1+P2+P3, P1, P2, P3). The binned rotation curves were obtained by adopting the best-fit position angle for the rotation axis (Table~\ref{bestfit}), assuming an equal number of stars in each bin (apart from the outermost bins) and using the Maximum Likelihood estimator of \citet{PM1993} to compute the mean line-of-sight velocity in each bin and its uncertainty. Again, note that the fits were performed on the discrete radial velocity measurements (taking into account uncertainties in the position angle of the rotation axis), not on the binned rotation curves. The 1-$\sigma$ envelopes on the best-fit rotation curves shown in the bottom panels of Figures \ref{rot_curves_P2_P3} and \ref{rot_curves_P1_P2} were obtained in an analogous way as described above for the dispersion profiles.
%Maybe fold binned rotation curve around symmetry axis to get more signal?

We first note from Table \ref{bestfit} that the mean velocity ($v_0$) of all the chemical subpopulations considered is consistent (within 1$\sigma$) with the mean velocity of the cluster as a whole. The line-of-sight velocity dispersion of the different chemical subsamples are also consistent with each other within 1$\sigma$, with only a slight hint of a larger dispersion for population P1 compared to population P2, and a larger dispersion for population P2 compared to population P3, a trend that is in keeping with the more centrally concentrated distribution of population P3 reported by \citet{JohnsonPilachowski2012} \citep[see e.g.][]{VHBetal2015}. The observed dispersion profiles decrease with increasing radius within the range of our data.

A clear rotation signature is found for the cluster as a whole, with an amplitude $A_{\rm rot}=2.7^{+0.9}_{-0.8}$~\kms. We emphasise again that the rotational amplitudes reported here are lower limits to the three-dimensional rotational amplitudes because our data is only sensitive to the line-of-sight component of rotation, and the inclination of the rotation axis with respect to the line of sight is unknown for M13. Rotation is also obvious in population P3 ($A_{\rm rot}=5.6^{+1.7}_{-1.6}$~\kms), while a much weaker rotational signal is found for the dominant population P2 ($A_{\rm rot}=1.3^{+1.1}_{-0.9}$~\kms \ but consistent with no rotation within 2$\sigma$), suggesting that population P3 is driving the rotation signature found in the full sample. Unsurprisingly, with a small number of stars in the P1 subsample, the kinematics of the ``normal" population is poorly constrained, in particular its rotational amplitude (weak rotation but also consistent with no rotation within 2$\sigma$) and the position angle of its putative rotation axis. For the cluster as a whole and all subsamples, the rotation curve is consistent with solid-body in the inner parts followed by a decline beyond $\sim2-3$~arcmin, although the peak at intermediate radii is only visible in the binned rotation curve of groups P1+P2+P3, P2, and to a lesser extent P3. This is in keeping with observations of the rotation curve of other clusters such as $\omega$~Cen \citep[e.g.][]{Merritt1997, MeylanMayor1986, VdV2006, Sollima2009} and 47 Tuc \citep[e.g.][]{MeylanMayor1986}. We note that for population P3, it is possible that we miss the declining part of the rotation curve because of the spatial sampling and centrally concentrated distribution of that population. Interestingly, the inferred position of the rotation axis of the different populations considered is in agreement within 1$\sigma$ uncertainties, as can be seen from Figure~\ref{vel_map} and Table~\ref{bestfit}.

The biggest kinematic differences are found when comparing the rotation curves of the P2 and P3 subsamples (Figure~\ref{rot_curves_P2_P3}), with very similar results obtained when comparing P3 with the rest of the cluster stars (P1+P2). With the kinematics of population P1 less well constrained, we cannot identify any difference between the rotation curve of this population and that of P2 or P3 (see for example Figure~\ref{rot_curves_P1_P2}). 

In Figure~\ref{post_Arot_P2_P3}, we illustrate the difference in the rotational amplitude of P2 and P3 by overlaying the posterior probability distribution of $A_{\rm rot}$ for these two populations. To quantify this further, we show in Figure~\ref{post_P2_P3} the posterior probability distribution of the difference in rotational amplitude between P3 and P2. This distribution peaks at $\sim4$~\kms \ and is clearly skewed towards P3 having a larger rotational amplitude than P2, with a 98.4\% probability that the rotation amplitude of P3 is larger.

To further illustrate the observed kinematic differences between P2, P3, and the cluster as a whole, we show in Figure~\ref{vrot_sigma} the radial profile of $V_{\rm rot}/\sigma_{\rm los}$ for these samples averaged along specific line of sights (with inclination angles of $i=90^{\circ}$, $i=60^{\circ}$, and $i=30^{\circ}$ between the rotation axis and the line of sight) in order to give an idea of the degeneracy between inclination angle and intrinsic rotation. Note that the inclination angle of the rotation axis of M13 is not constrained observationally, but $i=60^{\circ}$ was adopted as a rough estimate providing a satisfying agreement between observations and self-consistent rotating models by \citet{Lupton1987}. To build the $V_{\rm rot}/\sigma_{\rm los}$ profiles as a function of projected distance $R$ from the centre of the cluster, we average the absolute value of $V_{\rm rot}$ over all angles in the meridional plane (i.e. x-y plane in Fig.~\ref{vel_map}) for a given $R$ and then divide by $\sigma_{\rm los}(R)$. This is because for the assumed cylindrical rotation, $V_{\rm rot}$ was defined above as a function of the distance from the rotation axis, not as a function of distance from the cluster centre.

$V_{\rm rot}/\sigma_{\rm los}$ of the whole sample (P1+P2+P3) peaks at a value of $\sim0.4$ assuming $i=90^{\circ}$, but the peak value could be as high as $\sim0.7$ for a less favourable inclination of $i=30^{\circ}$. This peak occurs around 2 arcmin from the centre of the cluster, beyond which the $V_{\rm rot}/\sigma_{\rm los}$ profile is almost flat or possibly decreasing only very slightly within the range of our data. A similar flattening of the profile is found for subsamples P2 and P3, with a peak value reached further out ($\sim5$~arcmin) for population P3. It is interesting to note the importance of ordered motions for population P3, with $V_{\rm rot}/\sigma_{\rm los}$ reaching a maximum of at least 0.5 but possibly even larger than 1 or 2 when considering statistical uncertainties and the unconstrained inclination of the rotation axis. In comparison, the maximum $V_{\rm rot}/\sigma_{\rm los}$ for P2 is generally at most $\sim0.5$ or lower.

\begin{figure*}
\centering
\includegraphics[width=18cm]{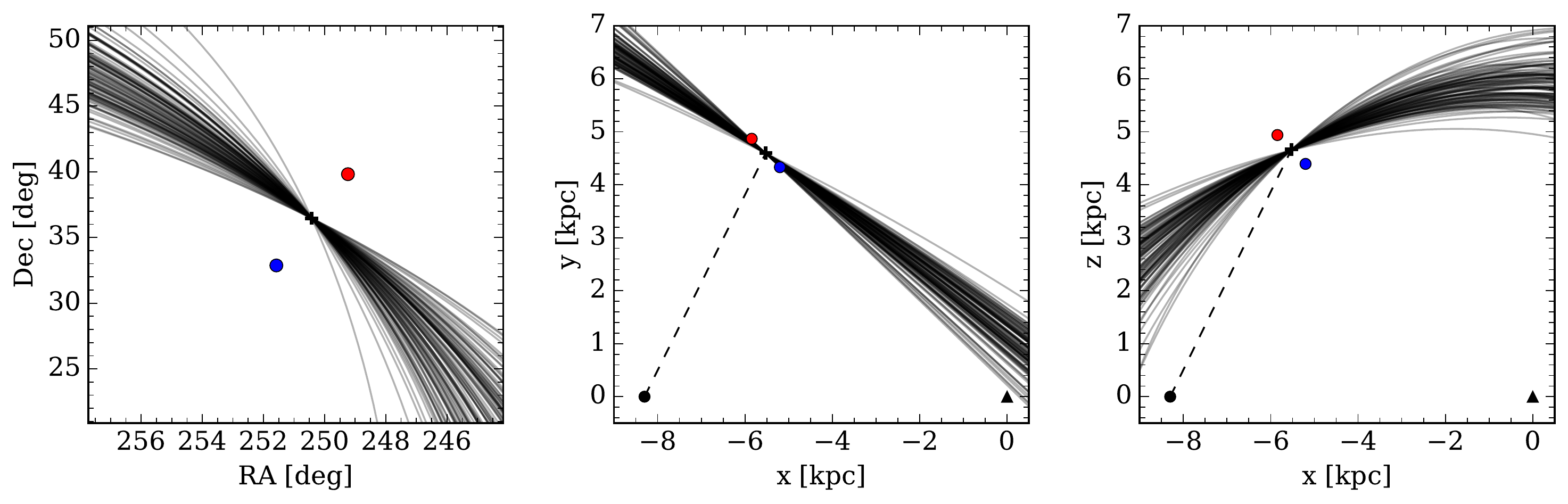}
\caption{Orbit realisations (solid black lines) in the Milky Way potential for the GC M13, with variations between the different realisations capturing the uncertainty in the measured velocity of the cluster. The orbits are shown in the plane of the sky in equatorial coordinates (left panel), and in the Galactocentric cartesian frame ($R_{\odot} = 8.3$~kpc; $y,z_{\odot}=0$) as a top view (x-y plane; middle panel) and side view (x-z plane; right panel). The position of the sun is indicated with a black circle, the Galactic centre with a black triangle, M13 with a black plus sign, the line of sight with a dashed line, and the blue and red circles mark the position of the Lagrange points, which have been offset from the cluster centre by 0.5 kpc in this schematic picture.}
\label{M13_orbit}
\end{figure*}

\section{Discussion}

\label{discussion}

\subsection{Possible sources of contamination to the rotation signature}
\label{contaminant}

Before discussing the interpretation of the rotational signature reported in the previous section, we consider here possible sources of contamination that could affect our observations and introduce a spurious rotational signal in our radial velocity dataset.

One of these is a possible alignment of tidal tails with the line of sight, which could mimic a gradient in the mean radial velocity across the cluster (otherwise attributed to rotation) if the leading and trailing tails are seen in projection on the cluster. Stars just escaping from the Lagrange point towards the Galactic centre (for the Lagrange point closest to the Galactic centre) or in the direction opposite to the Galactic centre (for the Lagrange point furthest from the Galactic centre) could also act as a source of contamination.

To verify possible line-of-sight alignments of the tidal tails or Lagrange points of M13, we investigate the orientation of the cluster's orbit and Lagrange points. The present day velocity and uncertainty were taken from \citet{Dinescu1999}. We perform 200 orbit integrations using galpy \citep{Bovy2015} in the \textsc{MWPotential2014} potential. Each orbit realisation assumes a velocity sampled from a normal distribution with mean and dispersion equal to the literature value and uncertainty. In Figure \ref{M13_orbit}, the black lines show the orientation of possible orbits for M13 in equatorial coordinates and also in a Galactocentric cartesian frame with $R_ {\odot}=8.3$~kpc, $y_{\odot}=z_{\odot}=0$ in top view (x-y) and side view (x-z). The blue (red) dot shows the orientation of the Lagrange point that is less (more) distant from the Galactic centre. These have been offset from the cluster centre by 0.5 kpc for illustration purposes. In this configuration, stars escaping from the less (more) distant Lagrange point would have radial velocities that are blueshifted (redshifted) with respect to the systemic radial velocity of the cluster. The position of the Sun, M13 and the Galactic centre are also indicated in Figure \ref{M13_orbit}, and the line of sight is shown with a dashed line.

From this analysis, we infer that there is no alignment with the line of sight from the tidal tails or the Lagrange points. We cannot discard that there is some overlap of escaper stars with the line of sight, but given the geometry of the orbit of M13, these would be mainly observed in the outskirts of the cluster. The fact that the rotation curve of M13 peaks at intermediate radii and then declines further out argues against the idea that these escapers in the outskirts of the cluster constitute a significant source of contamination to the observed rotational signature. Moreover, it is for the more centrally concentrated population \citep[P3;][]{JohnsonPilachowski2012} that we infer the fastest rotation, which also argues against a significant contribution to the rotation signal from escaper stars in the outskirts of the cluster, because we would otherwise expect the least centrally concentrated populations (P1 and P2) to be the most affected. Finally, given that stars escaping from clusters are preferentially lower mass stars, it is doubtful that a large population of escapers would contaminate our sample of giants, unless the cluster has lost a large fraction of its mass. The present-day mass function of M13 \citep[$\alpha=-0.98\pm0.02$ for $0.3 \ M_{\odot} \lesssim m \lesssim 0.8 \ M_{\odot}$, with $dN/dm \propto m^\alpha$;][]{Paust2010} is not strongly depleted in low-mass stars and is indeed consistent with the cluster having lost less than $\sim30\%$ of its initial mass due to two-body relaxation \citep{Trenti2010}, so we expect the mean mass of the escaper stars to be much lower than the typical mass of our observed giants. Tides can also give rise to rotation inside the cluster, because unbound stars on retrograde orbits (with respect to the direction of the cluster's orbit) are more stable against escape than prograde orbits \citep[e.g.][]{Tiongco2016}. However, the resulting rotation signature would be close to solid-body rotation and stronger in the outer parts of the cluster, contrary to our observations for the same reasons as outlined above.

Another effect could potentially introduce a spurious rotational signal, but it is expected to be most important in objects with a larger angular extent on the sky such as dwarf spheroidal satellite galaxies of the Milky Way. In these extended objects, the mean line-of-sight velocity of a star will depend on the position of the star because of geometrical projection effects due to the proper motion of the object, introducing a velocity gradient mimicking solid-body rotation \citep[e.g.][]{Battaglia2013}. Note again that the rotation curve of M13 appears to peak at intermediate radii and is not solid-body all the way to the outer parts of the cluster as would be expected if this geometrical effect was important. Moreover, the spurious solid-body rotation expected in this case would result in a velocity field with a velocity gradient parallel to the proper motion direction. From joint inspection of Figures~\ref{vel_map} and \ref{M13_orbit} (left panel), we can see that this is not the case and that the velocity gradient across the cluster is actually almost perpendicular to the orbit of the cluster projected on the sky. We thus conclude that the rotation signature detected is not the result of geometrical/projection effects.

\subsection{Global rotation of M13}

The clear global rotation signal that we detected in M13, with an amplitude of $A_{\rm rot}=2.7^{+0.9}_{-0.8}$~\kms, is consistent with the rotation curve obtained by \citet[][see their Fig. 5]{Lupton1987} from a similar sample size but with less precise radial velocity measurements. These authors also report a 1D central velocity dispersion in the giants of $\sim7$~\kms, consistent with our inferred central velocity dispersion of $\sigma_0=6.6^{+0.8}_{-0.6}$~\kms. From Integral Field Unit (IFU) observations, \citet{Fabricius2014} measured a central velocity gradient of $||\nabla v|| = 1.7\pm0.1\pm0.1$~km~s$^{-1}$~arcmin$^{-1}$ (attributed to cluster rotation) in the inner $\sim1-2$~arcmin of M13. This is in excellent agreement with   the central velocity gradient $||\nabla v|| \simeq 1.8$~km~s$^{-1}$~arcmin$^{-1}$ implied by the best-fit rotation curve of our full sample (where $||\nabla v|| =  \left. \delta (V_{\rm rot}  \sin{i}(X_{\rm PA_{0}}) )/\delta X_{\rm PA_{0}} \right\vert_{X_{\rm PA_{0}}=0} = 2 A_{\rm rot}/R_{\rm peak}$). Note that our spectroscopic data do not probe as far in as the IFU data of \citet{Fabricius2014}, but the good match between their observed central velocity gradient and the one implied by our best-fit rotation curve lends further support to our assumption of solid-body rotation in the inner parts of the cluster. The position angle of the rotation axis reported by \citet{Fabricius2014}, PA$_{0}=16.5\pm3.6\pm7.8^{\circ}$ \footnote{\citet{Fabricius2014} actually report PA$_{\rm kin}=106.5^{\circ}$, but in the convention adopted by these authors we would have PA$_{0}= {\rm PA}_{\rm kin}-90^{\circ} = 16.5^{\circ}$.} is also perfectly consistent with our inferred value of PA$_{0}=14^{+19}_{-16}$$^{\circ}$.   

M13 not only rotates, it is also somewhat oblate\footnote{\citet{Kadla1976} also suggest that the cluster may be slightly prolate in the very central parts (radii less than 0.7 arcmin), but these results should be considered uncertain due to the small number of stars contributing to the light in these regions \citep[see discussions in][]{Lupton1987}.}. This flattening was first observed using Palomar Sky Survey prints by \citet{Kadla1976}, who presented a curve of flattening and position angle of the semi-major axis as a function of distance from the centre. They found that the axial ratio of isodensity contours falls from around 1.0 at small radii ($\sim0.7$~arcmin) to about 0.85 at 6~arcmin, with the position angle of the semi-major axis remaining roughly constant at $\sim110-120^{\circ}$. \citet{Fabricius2014} also derived the central ellipticity and position angle for the best-fitting ellipsoid using published catalogues of the ACS survey of GCs. In the inner 100 arcseconds, they measured an ellipticity of $\sim0.02$ (i.e. axial ratio of $\sim0.98$) with a position angle for the semi-major axis of $115.8\pm3.8\pm16.8^{\circ}$. These position angles for the major axis of the photometric ellipsoid are consistent with being shifted by $90^{\circ}$ from the position angle of the rotation axis, in agreement with what is expected if internal rotation is responsible for the flattening of the cluster\footnote{For detailed examples of the connection between rotation and flattening in the GCs $\omega$~Cen and 47 Tuc, see \citet{Bianchini2013}.}. This idea is further supported by the rotating distribution-function based models of M13 by \citet{Lupton1987}, which show reasonable agreement with the observed variation of the axial ratio of isophotes as a function of the semi-major axis \citep[based on the data by][]{Kadla1976}. The best-fit models of M13 from \citet{Lupton1987} also have a position angle of $130\pm15^{\circ}$ for the semi-major axis, which is consistent with the measurements referenced above. However, note that other measurements of the position angle of the photometric semi-major axis appear discrepant from the ones listed above. \citet{WS87} and \citet{CC10} found a position angle for the semi-major axis of 52 and 162$^{\circ}$, respectively, with approaches sensitive to the inner \citep{WS87} and outer parts of GCs \citep{CC10}. The two studies report an axial ratio of 0.89 and 0.88, respectively. Note that the \citet{WS87} work is based on optical data, while \citet{CC10} is based on the spatial distribution of 2MASS point sources (i.e. infrared), and differences in the position angle from these two studies are frequent.

M13 is only one example in a long and growing list of GCs showing evidence of internal rotation and/or deviations from sphericity \citep[e.g.][]{Geyer1983, MeylanMayor1986, WS87, G94, PC1994, MH1997, VLLP2002, AndersonKing2003, CC10, Lane2010, Bellazzini2012, Bianchini2013, Fabricius2014, Kacharov2014, Lardo2015}. That said, the link between rotation and flattening is not necessarily obvious, because other factors such as pressure anisotropy and Galactic tides can lead to flattening  \citep[e.g.][]{Stephens2006, vdB2008}. A simple and commonly used tool to estimate the importance of rotation in shaping the morphology of GCs is to place them in the $V^{\rm rot}_{\rm max}/\sigma_0$ vs. $\epsilon$ diagram, where $V^{\rm rot}_{\rm max}/\sigma_0$ is the ratio of the observed maximum of the line-of-sight rotation profile to the central line-of-sight velocity dispersion and $\epsilon$ is the (global) ellipticity. We note that this approach is no substitute to detailed dynamical modelling, because it only compares global quantities while rotation and ellipticity can vary significantly with radius as the anisotropy parameter changes, and the position of a cluster in the $V^{\rm rot}_{\rm max}/\sigma_0$ vs. $\epsilon$ diagram is also sensitive to inclination effects \citep[see discussion in][]{Bianchini2013}. The position of a cluster in the diagram is often compared to the relation between $V^{\rm rot}_{\rm max}/\sigma_0$ and $\epsilon$ expected for oblate rotators viewed ``edge-on" \citep{Binney2005, Cap2007}. Assuming an average ellipticity of $\sim0.9$ for M13 (see above), and estimating $V^{\rm rot}_{\rm max}/\sigma_0 \simeq 0.4\pm0.1$ from the values of Table~\ref{bestfit} for the whole sample (and bearing in mind inclination effects), we find that M13 is in good agreement with the expected relation for an isotropic oblate rotator \citep[comparing for example with figures from ][]{Bianchini2013, Kacharov2014}. This suggests that the flattening is consistent with being caused by internal rotation. A value of $V^{\rm rot}_{\rm max}/\sigma_0 \simeq 0.4\pm0.1$ also places M13 among the GCs having the largest fraction of energy in rotation. This evidence and the remaining discrepancies discussed above (e.g. mismatches between the position of the semi-major axis from different studies) call for a more detailed study modelling the interplay of rotation, pressure anisotropy, Galactic tides and flattening in this cluster, which will be helped by the precise radial velocities reported here and future proper motions from the Gaia satellite. This is however beyond the scope of this paper.

Another useful diagnostic is the ratio of rotational amplitude to velocity dispersion as a function of radius (Figure~\ref{vrot_sigma}). Based on evolutionary 2D Fokker-Planck models of rotating clusters, we expect the $V_{\rm rot}/\sigma$ profile to reach a maximum between about one and several half-mass radii and then decrease with increasing radius \citep{Kim2002, Fiestas2006}. Such a profile has been observed in other clusters like $\omega$~Cen and 47 Tuc \citep{MeylanMayor1986}, and is qualitatively consistent with what we observe in M13\footnote{When comparing our $V_{\rm rot}/\sigma$ profile with those of \citet{Fiestas2006} and \citet{Kim2002}, note that our $V_{\rm rot}/\sigma$ values may be underestimated because of the way we define $V_{\rm rot}$ compared to these authors. Our understanding is that they represent $V_{\rm rot}$ as the azimuthal velocity evaluated on the equatorial plane, while we average the rotational velocity over all angles in the meridional plane. With our assumed cylindrical rotation, this yields lower values of $V_{\rm rot}$ at a given radius than if we reported the rotational velocity on the equatorial plane.}. As a cluster evolves it loses mass and angular momentum, and the radial position of the peak of the $V_{\rm rot}/\sigma$ profile will move outward. The maximum value of $V_{\rm rot}/\sigma$ at a given time also decreases monotonically with the dynamical age of the cluster \citep{Kim2002}. Together, these features can provide useful constraints on evolutionary models and may allow to put limits on the initial amount of rotational energy in clusters. For example, the relatively flat $V_{\rm rot}/\sigma$ profile of M13 over several half-mass radii along with its modest maximum $V_{\rm rot}/\sigma$ value (even when considering more favourable inclination angles) could be indicative of a post-core-collapse stage for this cluster, but we note that is rather speculative and based solely on a qualitative comparison with the results of  \citet[][see their Fig. 16]{Kim2002}. Detailed comparison with evolutionary models will be necessary to quantify this further.

\subsection{Connection between global rotation and multiple populations}

There may exist important connections between rotation and the formation of GCs and their multiple populations. While today the amount of rotational energy in clusters is typically not dominant, it is also not negligible \citep[$0< V^{\rm rot}_{\rm max}/\sigma_0 <  0.5$, e.g.][]{MH1997}, and it could have been much more important initially because rotation is progressively wiped out as clusters evolve dynamically. Rotation is a natural consequence of the cluster formation process from the collapse of a star-forming cloud with net angular momentum, and differential rotation would naturally arise in star clusters that form by undergoing the process of violent relaxation in the tidal field of their host galaxy \citep{Vesperini2014}. Significant rotation at birth may be a ubiquitous property of GCs, and therefore rotation could be a key ingredient for understanding the widespread phenomenon of multiple populations in GCs \citep[e.g.][]{Bekki2010, VHBetal2015, Mastrobuono2016}.

Based on a sample of 25 Galatic GCs, \citet{Bellazzini2012} reported possible correlations between global rotation (specifially $A_{\rm rot}$ and $A_{\rm rot}/\sigma_0$, where $A_{\rm rot}$ is a proxy of the mean rotational amplitude) and other cluster properties, some of which are sensitive to the characteristics of multiple populations in GCs.

They suggested that there may be a weak anticorrelation between $A_{\rm rot}/\sigma_0$ and the inter quartile range (IQR) of the Na-O anticorrelation (a parameter defined by \citet{Carretta2006} to measure the extension of the Na-O distribution in clusters and thus directly related to the properties of multiple populations). The suggested trend is such that clusters having a larger $A_{\rm rot}/\sigma_0$ show a less extended Na-O anticorrelation. With a large $V_{\rm rot}/\sigma_0$ and a very extended Na-O anticorrelation, M13 does not support this suggestion, and is instead similar to other outliers of such a possible trend like the peculiar NGC 2808 and $\omega$~Cen.

A somewhat clearer trend was reported by \citet{Bellazzini2012} between $A_{\rm rot}/\sigma_0$ and metallicity ([Fe/H]), with higher metallicity clusters having typically larger values of $A_{\rm rot}/\sigma_0$. M13 does not really strengthen that conclusion with its high $A_{\rm rot}/\sigma_0$ compared to all other clusters in the moderately metal-poor regime around ${\rm [Fe/H]}=-1.5$.

Another correlation was found by \citet{Bellazzini2012} between $A_{\rm rot}/\sigma_0$ and the horizontal branch morphology parameter ($\frac{B-R}{B+R+V}$), with GCs having a bluer horizontal branch typically displaying lower $A_{\rm rot}/\sigma_0$. Metallicity is the most important parameter in shaping the morphology of the horizontal branch \citep[e.g.][]{Lee1990} so it might play a role here as well, but since this correlation with the horizontal branch morphology parameter was found to be stronger than the correlation between $A_{\rm rot}/\sigma_0$ and [Fe/H], there are likely additional parameters at play. Possible additional parameters include age and/or chemical abundance differences (in particular He) linked to variations in the properties of multiple populations \citep[e.g.][]{VandenBerg2013}. In any case, M13 is a clear outlier that does not fit this picture with a large $A_{\rm rot}/\sigma_0$ despite its bimodal and extremely blue horizontal branch with $\frac{B-R}{B+R+V}=0.976$ \citep{Sand2010}.

Finally, note that \citet{Bellazzini2012} also found significant correlations between the amplitude of rotation $A_{\rm rot}$ and some intrinsic cluster properties (namely $\frac{B-R}{B+R+V}$, $M_V$, $\sigma_0$, and [Fe/H]; with $M_V$ and $\sigma_0$ both tracing cluster mass). The correlation of $A_{\rm rot}$ with the horizontal branch morphology parameter was found to be particularly strong, but even stronger and more significant was the correlation of $A_{\rm rot}$ with two linear combinations of parameters: $\frac{B-R}{B+R+V}+0.47 M_V$ and $M_V - 1.73$ [Fe/H]. With $M_V$ found to correlate with the level of He enrichment \citep[brighter clusters typically have larger He abundance spreads;][]{Milone2015} and also found to play a role in bi-variate correlations involving parameters related to the anticorrelation phenomenon \citep{Carretta2009}, all these possible relations hint at a complex interplay between cluster mass, rotation, and the chemical properties of multiple populations. Investigating the kinematics of subpopulations in massive and rapidly rotating GCs with distinct or extreme chemical patterns like M13 may help to unveil some of the details behind of this interplay.

\subsection{Difference in the rotational amplitude of multiple populations: implications for formation scenarios}

The multiple populations of M13 are not fully mixed \citep[the extremely O-depleted population is more centrally concentrated;][]{JohnsonPilachowski2012}, as expected based on its relatively large mass of $\sim6\times10^5 \ \msun$ \citep{Leonard1992}, its Galactocentric radius of 8.4~kpc \citep{Harris1996}, and a mass function that is not strongly depleted in low-mass stars suggesting only moderate mass loss due to two-body relaxation (see discussion and references in Section~\ref{contaminant}). On that basis, it is also not surprising that we were able detect kinematic differences between subpopulations \citep{VHBetal2015}.

But before we can establish that the observed difference in rotational amplitude between the multiple populations of M13 stems for the formation and early evolution of the cluster, we first need to consider whether this signature could simply be a product of dynamical evolution. This could happen if the distinct chemical populations also have different mean stellar masses. One way to achieve this is through differences in their initial He fraction. Using far-ultraviolet and optical CMDs from HST, \citet{Dalessandro2013} estimated a maximum helium abundance of ${\rm Y}\sim0.3$ for stars in M13 \citep[see also][]{VandenBerg2013}. Based on Dartmouth isochrones with $\rm{[Fe/H]}=-1.5$, an age of 12.25 Gyr and $[\alpha/{\rm Fe}]=0.4$ \citep{VandenBerg2014} appropriate for M13, giants with normal He ($\rm{Y=0.247}$) have masses of $0.82$~$\msun$, while He-rich giants ($\rm{Y}=0.33$) have masses of $0.7$~M$_{\odot}$. Assuming that the extremely O-depleted giants in M13 are also the most He-enriched, we estimate that they would be about $0.1$~$\msun$ less massive than their He-normal counterparts.

The dynamical evolution of rotating clusters with two mass components or a mass spectrum has been studied with Fokker-Planck models \citep{Kim2004}, direct $N$-body simulations \citep{Ernst2007}, or both \citep{Hong2013}. From these studies, the main dynamical evolution process that would lead to a difference in rotational amplitude between different mass species actually causes the most massive stars in the central parts to rotate faster than lower mass stars. Two main factors are instrumental in determining the evolutionary behaviour of the rotational properties of the models: the transport of angular momentum within the radial extension of the system, as driven by collisional relaxation (which obviously takes place also in equal-mass systems) and the exchange of angular momentum between different mass species, in association with the tendency towards energy equipartition. The second factor is intrinsically linked to the process of mass segregation, i.e. the fact that more massive stars tend to become progressively more radially concentrated, as a result of their relaxation-driven evolution. In the presence of global internal rotation, the interplay between mass segregation and transport of angular momentum is highly non-trivial, but the investigations quoted above have already offered some evidence that heavier component becomes more centrally concentrated and are characterised by a higher angular velocity, compared to the lower mass component, which is spatially more diffuse and slower rotating.  The two components then experience a rather different evolution, with the angular velocity of the heavier component increasing very rapidly during the early phases, while that of the lighter stars shows only a moderate increment \citep[see Fig. 6 from][]{Kim2004}. In addition, these results seem to have a significant dependence on the initial rotation strength of the systems \citep[see also][]{Hong2013}, and it should be stressed that, in this context, the different mass components start with the same value of angular velocity.

%For example, based on their Fokker-Planck models, \citet{Kim2004} argued that as long as dynamical friction dominates in the competition with angular momentum exchange, the more massive stars lose random energy and angular momentum and sink towards the centre of the cluster, but their remaining angular momentum is sufficient to speed them up rotationally. The same behaviour was observed in the $N$-body models of \citet{Hong2013}.

From the mass differences estimated above for the extremely O-depleted (He-rich) giants compared to He-normal giants, the expected effect of dynamical evolution outlined above for multi-mass rotating systems is thus opposite to the difference in rotational amplitude that we detected in M13 (where the presumably less massive stars are rotating more rapidly). Moreover, differences in the peak rotational velocity of different mass species remain relatively small in the models of \citet{Kim2004} and \citet{Hong2013}, despite much larger mass ratios between the different components, while we find that the peak rotational velocity of the extreme population is much larger (a factor $\sim4$) than the peak rotational velocity of the other subpopulations. We thus argue that the faster rotation inferred for the extreme population (P3) in M13 is not, as far as we can tell, an effect produced by the long-term dynamical evolution of the cluster.
% Sugimoto & Makino 1989:"Two-body relaxation within the core now transports angular momentum outward, as discussed by Akiyama and Sugimoto (1989): The cluster as a self-gravitating system has, effectively, a negative moment of inertia, so that the decrease of specific angular momentum in the core results in an increase in the angular velocity. This intensifies the gradient of angular speed, which results in further transport of specific angular momentum outward. This process is the gravogyro instability (Hachisu 1979, 1982, Akiyama and Sugimoto 1989). It is quite analogous to the better known gravothermal casastrophe which is drive by an effective negative specific heat in the self-gravitating system. As a result of the gravogyro instability the angular momentum of the cluster is gradually squeezed out."

In a different context than multiple populations or mass species, \citet{Kim2002} mentioned another effect which could give rise to strong rotation for stars in the central regions of a cluster: collisions between stars or binaries could produce such a result since the collisions dissipate the relative kinetic energy of the colliding particles but conserve their net angular momentum in the cluster potential. If somehow the extreme O depletion and Na enrichment of population P3 can be linked to the formation of these stars via stellar mergers in the central regions of the cluster, in particular during its early evolution when the cluster was much denser, then this would potentially explain the more centrally concentrated distribution of this population and its faster rotation. In such a scenario, we would expect the rotation axis of this extreme population to align with the rotation axis of the bulk of the cluster stars, which is consistent with what we observe in M13.

Through conservation of angular momentum, a faster rotation and more centrally concentrated distribution is also expected for a stellar population that would form from dissipative accretion of enriched gas onto the central regions of the cluster \citep[e.g.][]{SF87}. This is what \citet{Bekki2010} also predicted in the context of self-enrichment of the cluster by AGB stars \citep[e.g.][]{DErcole2008}. In Section~\ref{Intro}, we pointed out some of the problems afflicting such a scenario, in particular the mass budget problem. The mass budget problem would perhaps be less of a concern if only the extreme population was formed in this way, but it would not solve many of the other problems with AGB ejecta as a source of pollution, for example the absence of gas, age spreads, or ongoing star formation in young massive clusters{\bf \footnote{This is assuming that young massive clusters are the younger counterparts of old, Galactic GCs. However, while they have sizes, luminosities, and masses comparable to those of old globulars, no sign of the presence of chemical anomalies has been detected so far in young massive clusters. On the other hand, the proposed scenarios for the origin of multiple populations typically do not make a distinction about the epoch of cluster formation and implicitly assume that the multiple population phenomenon should also be at play in young massive clusters.}}, and the cluster-to-cluster differences in the relative He and Na enrichment \citep{Bastian2015, Bastianetal2015}. Plus, it would not address the question of how the rest of the Na-enriched stars (with less extreme O depletion) got polluted. In this scenario, we would also expect the rotation axis to be the same for different subpopulations.

The opposite difference in rotation trend is expected from a scenario in which the most enriched stars are the ones initially crossing the core of the cluster and preferentially on radial (low angular momentum) orbits \citep[e.g. the early disc accretion scenario of][]{Bastian2013}. In this case, the rotational amplitude of the enriched stars at a given radius would be lower initially \citep[assuming the cluster has some net rotation; see][]{VHBetal2015}, which is contradicted by our observations of the faster rotation of population P3 in M13. As mentioned in Section~\ref{Intro}, this scenario is already disfavoured for other reasons.

\section{Summary and conclusions}
\label{conclusions}

We used radial velocities from spectra obtained with the WIYN telescope to probe the presence of kinematical differences among M\,13 multiple populations. The existing chemical tagging (in particular Na and O abundances) for our sample of stars coupled with the radial velocities measured in the present work make this dataset well-suited for this exploration.

We fitted a six-parameter kinematic model including rotation to the full dataset and various chemical subsamples. In addition to improved constraints on the global rotation and importance of rotational support in M13, we found a significant difference in the rotational amplitude of different subpopulations. The most likely difference between the rotational amplitude of the extremely O-depleted population and the rest of the sample was found to be $\sim4$~\kms (with the extreme population rotating faster), with  98.4\% probability that the rotational amplitude of this extreme population is larger than the rotational amplitude of the rest of the sample. We also find that the inferred position angle of the rotation axis for the different subpopulations considered is consistent, within uncertainties, with the hypothesis that all subpopulations share the same rotation axis, although it must be said that the rotation axis of the ``normal" halo-like subpopulation is poorly constrained.

We suggested that the origin of such a kinematical signature in M13 is likely a signature imprinted early in the formation history of this non-fully mixed GC. This would open up the possibility to use kinematics to distinguish between different scenarios for the formation of multiple populations \citep[e.g.][]{VHBetal2015}. We also argued, based on previous modelling that explored the evolution of multi-mass rotating systems, that this signature is unlikely to result from the long-term dynamical evolution of the cluster and be caused by mean stellar mass differences between the subpopulations.

Given the modest size of our sample, it will be desirable to follow up on these findings and hopefully improve the statistical significance of the reported difference in the rotation signals of subpopulations. Building a larger sample (in particular increasing the sample of O-depleted stars, which exhibit the stronger rotational amplitude in our dataset) would benefit from a different observational strategy, since chemical tagging using high-resolution spectroscopy is observationally expensive. An effective alternative method for tracing light-element abundance variations in GCs is based on photometry with narrow-band filters sensitive to C, N, and O abundances. This is the approach adopted by the Hubble Space Telescope UV Legacy Survey of Galactic GCs \citep{Piotto2015}, from which eventual catalogues of chemically tagged stars (which will include M13), complemented by Gaia proper motions and/or ground-based radial velocities, will facilitate a robust statistical exploration of the kinematical properties of Galactic GCs and their multiple populations.

Whether different degrees of rotation between chemical subpopulations in GCs is the norm or the exception remains to be studied on large samples of stars in more clusters. Similarly, whether the most O-depleted and Na-enriched stars systematically rotate faster than other subpopulations should be checked. Searching for these kinematic imprints and understanding why kinematic differences exist are crucial and necessary aspects of the study of GC formation and evolution.

\section*{Acknowledgements}
We thank the anonymous referee for constructive comments. We would also like to thank Mark Gieles, Nate Bastian, and Nikolay Kacharov for useful discussions and/or comments on the manuscript. M. J. C. and V. H.-B. acknowledge the hospitality of the Lorentz Center during the early stages of this collaboration. M. J. C gratefully acknowledges support from the Sonderforschungsbereich SFB 881 "The Milky Way System" (subproject A8) of the German Research Foundation (DFG). V. H.-B. acknowledges support from the Radboud Excellence Initiative Fellowship. C. A. P. gratefully acknowledges the support of the Kirkwood Research Fund at Indiana University. E. B. acknowledges financial support from the European Research Council (ERC-StG-335936, CLUSTERS). C. I. J. gratefully acknowledges support from the Clay Fellowship, administered by the Smithsonian Astrophysical Observatory. A. L. V. acknowledges support from the EU Horizon 2020 program (MSCA-IF-EF-RI 658088).

\bibliographystyle{mn2e}

\appendix
\section{Supplementary figures}
\label{appendixA}

For completeness, we show here one and two-dimensional projections of the posterior probability distribution of the six free parameters of our kinematic model for the different subsample considered.

\begin{figure*}
\centering
\includegraphics[width=\textwidth]{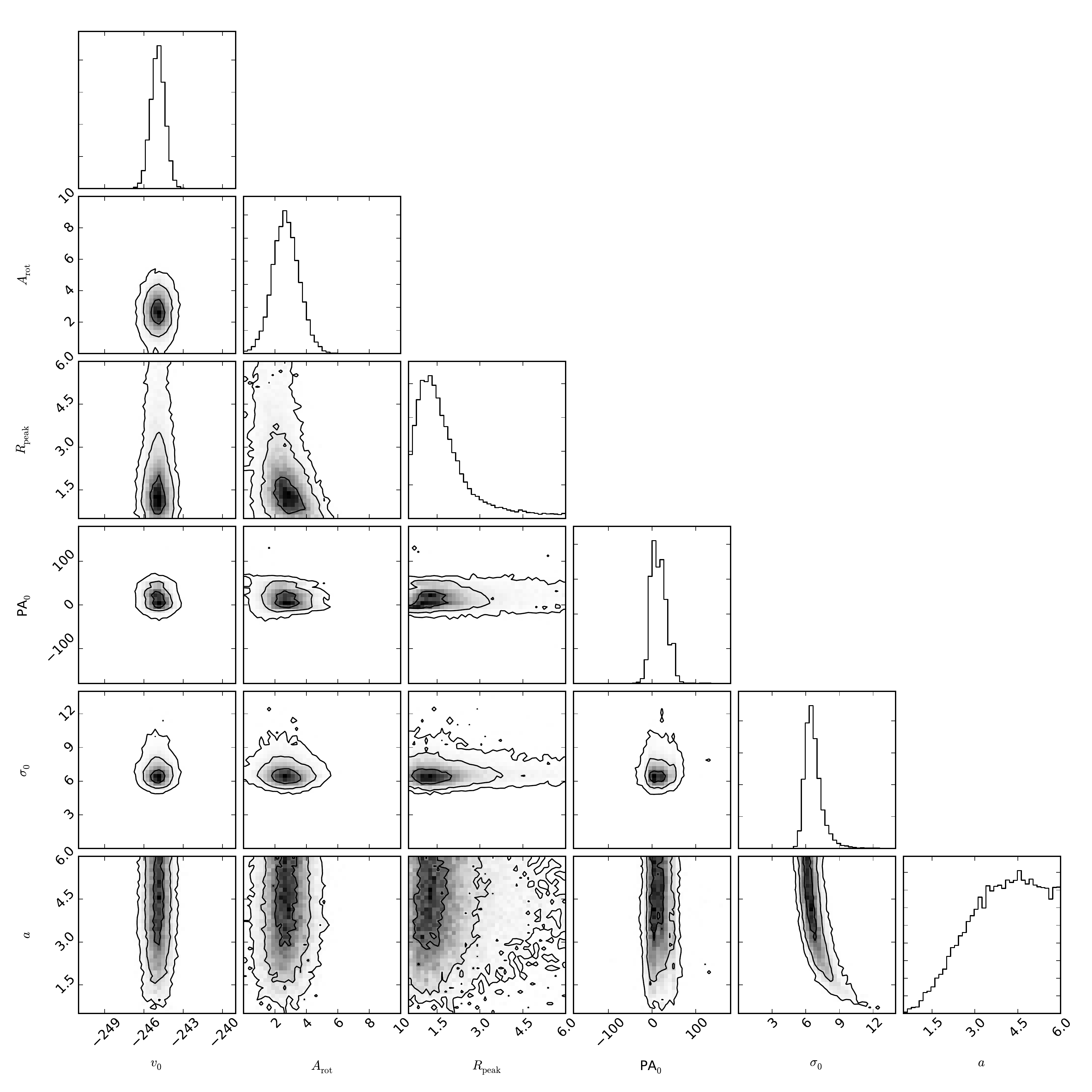}
\caption{Projections of the posterior probability distribution on the planes determined by every pair of parameters (see Section \ref{kinematics}) of our kinematic analysis of the full sample of stars in M13 (P1+P2+P3). Contours indicate 1, 2 and 3-$\sigma$ levels. Histograms representing the marginalised posterior probability distribution of each parameter are also shown.}
\label{post_all}
\end{figure*}

\begin{figure*}
\centering
\includegraphics[width=\textwidth]{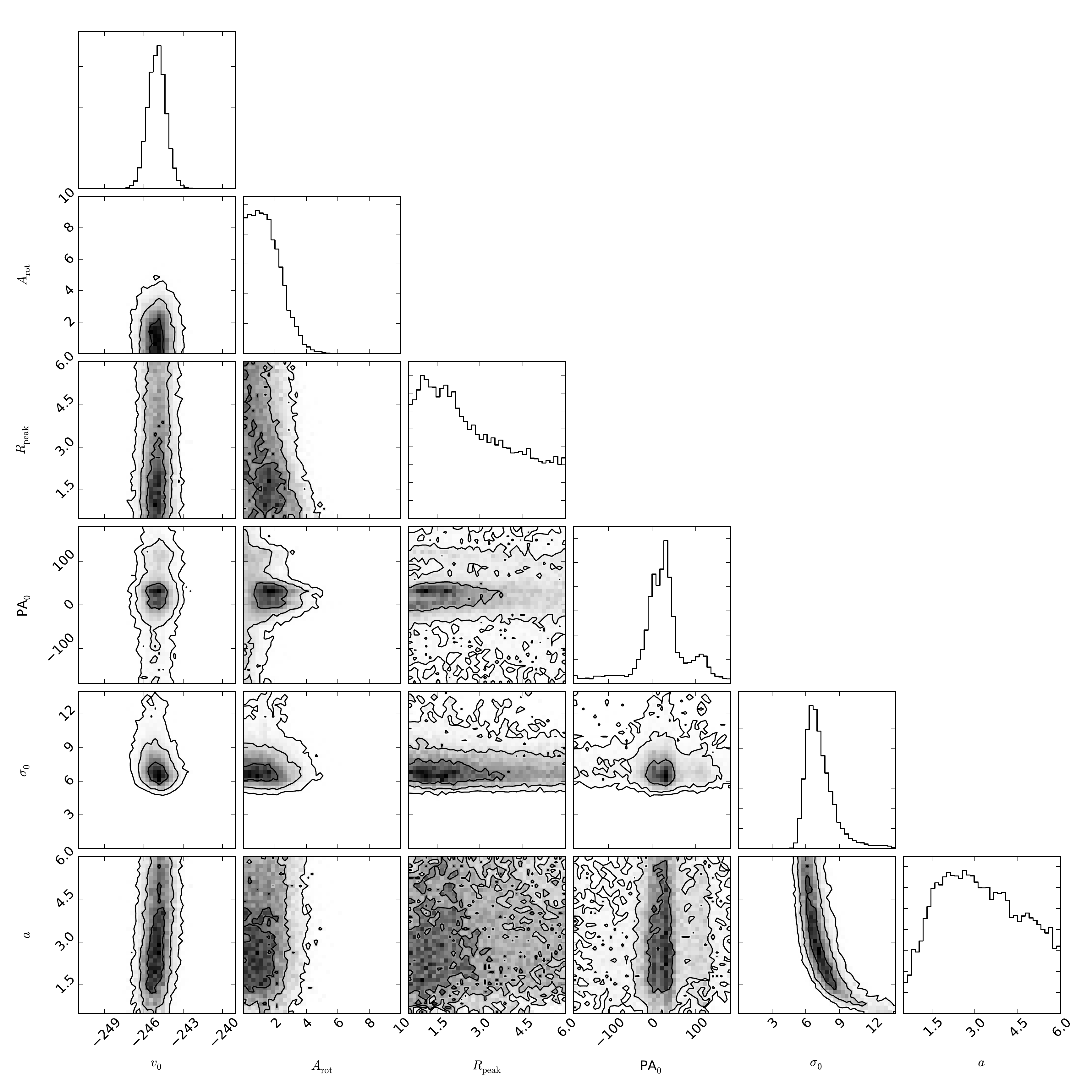}
\caption{Same as Figure \ref{post_all}, but for the subsample of ``intermediate" stars (P2).}
\label{post_intermediate}
\end{figure*}

\begin{figure*}
\centering
\includegraphics[width=\textwidth]{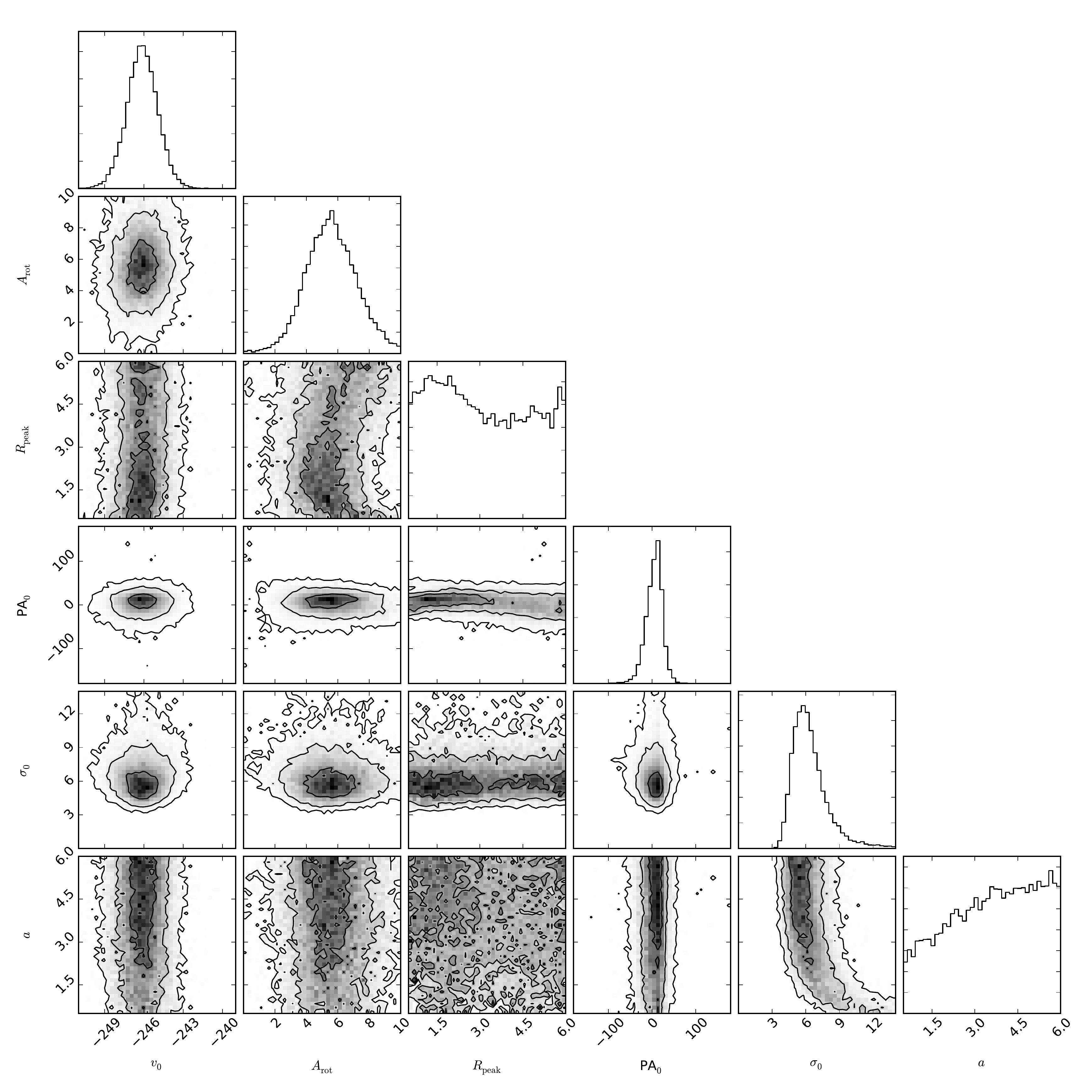}
\caption{Same as Figure \ref{post_all}, but for the subsample of ``extreme" stars (P3).}
 \label{post_extreme}
\end{figure*}
\newpage

\begin{figure*}
\centering
\includegraphics[width=\textwidth]{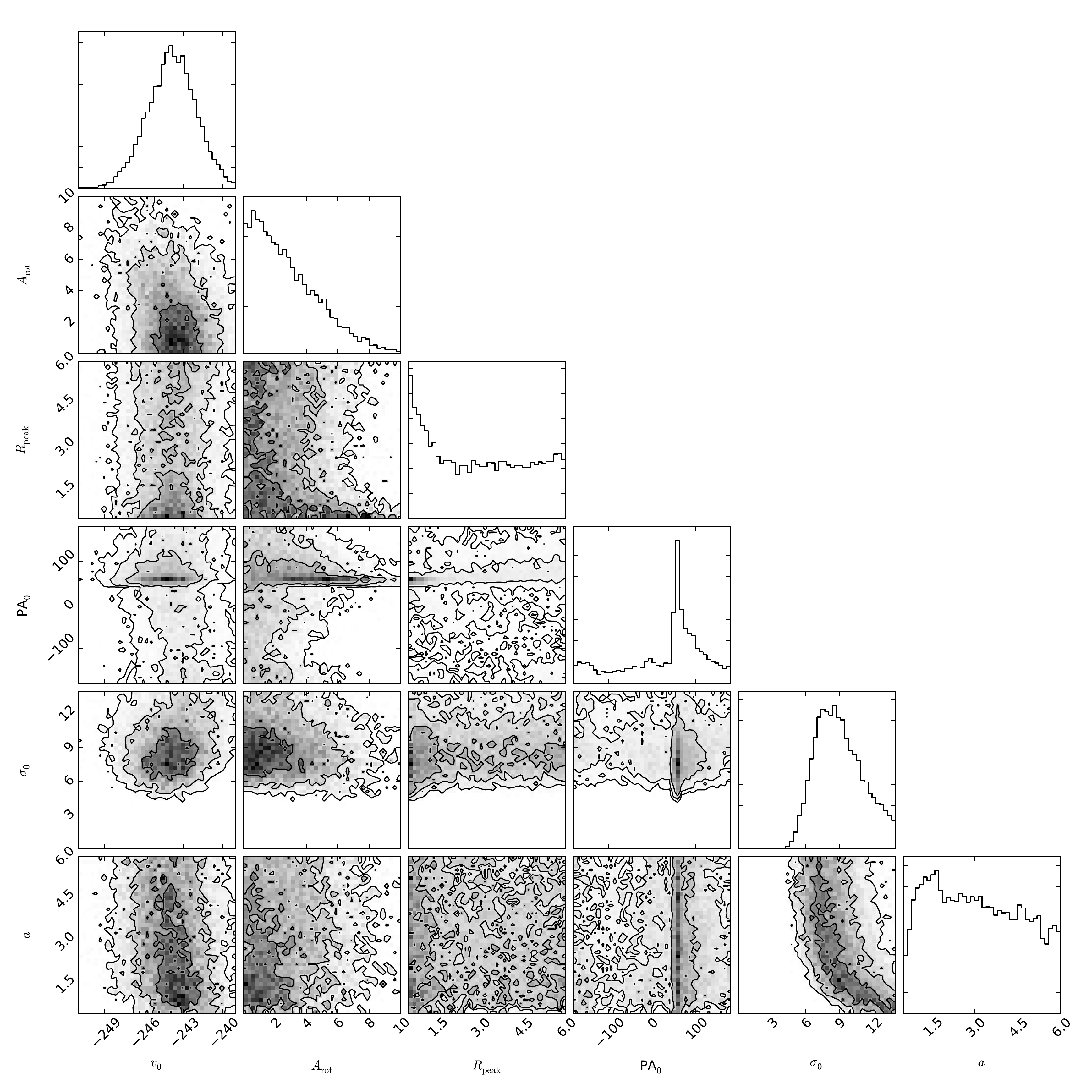}
\caption{Same as Figure \ref{post_all}, but for the subsample of ``normal" stars (P1).}
 \label{post_primordial}
\end{figure*}
\newpage

\begin{figure*}
\centering
\includegraphics[width=\textwidth]{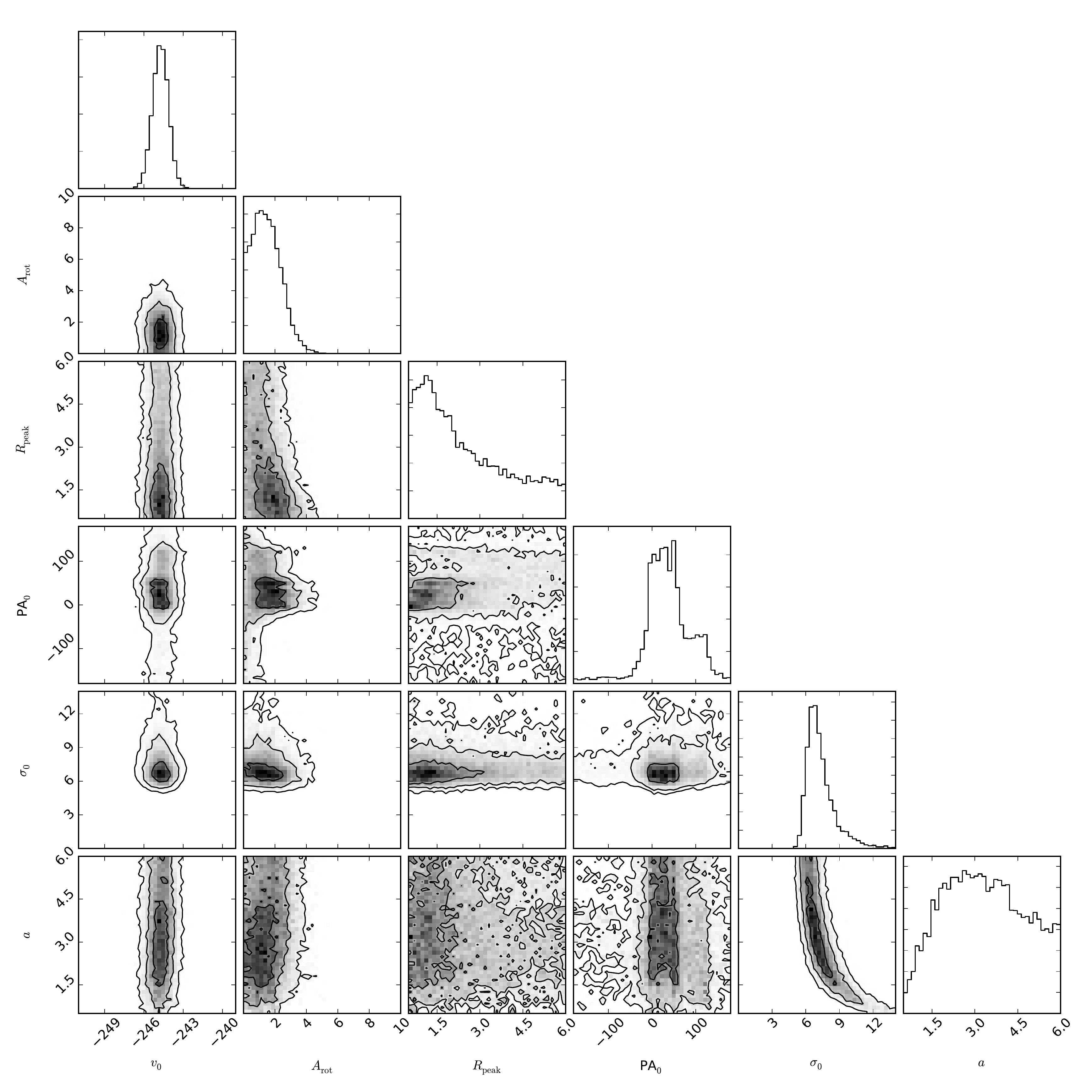}
\caption{Same as Figure \ref{post_all}, but for the subsample of ``normal" and ``intermediate" stars (P1+P2).}
\label{post_intprim}
\end{figure*}

\label{lastpage}

\end{document}